\documentclass[12pt]{article}  
\usepackage{amssymb,graphicx} 
\usepackage {apalike}
\begin{document}

\title{Space-time Philosophy Reconstructed \emph{via} Massive Nordstr\"{o}m Scalar Gravities?  Laws \emph{vs.} Geometry, Conventionality, and Underdetermination} 

\maketitle

\begin{center}
\author{J. Brian Pitts\\Faculty of Philosophy, University of Cambridge\\John Templeton Foundation grant \#38761\\jbp25@cam.ac.uk}
\end{center}

\pagebreak

\begin{abstract}

What if gravity satisfied the Klein-Gordon equation?  Both particle physics from the 1920s-30s and the 1890s Neumann-Seeliger modification of Newtonian gravity with exponential decay suggest considering a ``graviton mass term'' for gravity, which is \emph{algebraic} in the potential.  Unlike Nordstr\"{o}m's ``massless'' theory, massive scalar gravity is strictly special relativistic in the sense of being invariant under the Poincar\'{e} group but not the 15-parameter Bateman-Cunningham conformal group. It therefore exhibits the whole of Minkowski space-time structure, albeit only indirectly concerning volumes.  Massive scalar gravity is plausible in terms of relativistic field theory, while violating most interesting  versions of Einstein's principles of general covariance, general relativity, equivalence, and Mach.  Geometry is a poor guide to understanding massive scalar gravity(s):  matter sees a conformally flat metric due to universal coupling, but gravity also sees the rest of the flat metric (barely or on long distances) in the mass term.  What is the `true' geometry, one might wonder, in line with Poincar\'{e}'s modal conventionality argument?  Infinitely many theories exhibit this bimetric `geometry,' all with the total stress-energy's trace  as  source; thus geometry does not explain the field equations.  The irrelevance of the Ehlers-Pirani-Schild construction to a critique of conventionalism becomes evident when multi-geometry theories are contemplated.  Much as Seeliger envisaged, the smooth massless limit indicates underdetermination of theories by data between massless and massive scalar gravities---indeed an unconceived alternative.  At least one version easily could have been developed before General Relativity; it then would have motivated thinking of Einstein's equations along the lines of Einstein's newly re-appreciated ``physical strategy'' and particle physics and would have suggested a rivalry from massive spin $2$ variants of General Relativity (massless spin $2$, Pauli and Fierz found in 1939).  The Putnam-Gr\"{u}nbaum debate on conventionality is revisited with an emphasis on the broad modal scope of conventionalist views.  Massive scalar gravity thus contributes to a historically plausible rational reconstruction of much of 20th-21st century space-time philosophy in the light of particle physics.  An appendix reconsiders the Malament-Weatherall-Manchak conformal restriction of conventionality and constructs the `universal force' influencing the causal structure. 

Subsequent works will discuss how massive gravity could have provided a template for a more Kant-friendly space-time theory that would have blocked Moritz Schlick's supposed refutation of synthetic \emph{a priori} knowledge, and how Einstein's false analogy between the Neumann-Seeliger-Einstein modification of Newtonian gravity and the cosmological constant $\Lambda$ generated lasting confusion that obscured massive gravity as a conceptual possibility.  

\end{abstract}  %

keywords:  
Klein-Gordon equation;
Hugo von Seeliger;
Carl Neumann;
Bateman-Cunningham conformal group;
Nordstrom scalar gravity;
cosmological constant;
conventionalism;
underdetermination;
explanation and geometry;
massive gravitons;
neo-Kantianism

\tableofcontents



\section{Introduction}

Plausibly, when one comes to recognize the historical contingency of hitherto apparently unavoidable ideas about the world, one can take a more critical attitude and  rework one's beliefs to fit evidence and argument more fully.  Mach's historical-critical investigations in physics exemplified that idea.  More formally, it is plausible that the order in which one receives pieces of evidence ought not to affect one's final degrees of belief \cite{WagnerCommute}, a criterion for avoiding one sort of historical accident.  Failure of imagination can lead to our not entertaining theories that are comparably good to the ones that we did entertain; such unconceived alternatives undermine scientific realism \cite{StanfordUnconceived}.  In the interest of freeing ourselves from historical accidents regarding space-time theory, it is prudent, therefore, to employ whatever systematic means exist for generating plausible alternative theories. 

 Fortunately, there is a largely untapped source here, namely, the literature that studies all possible classical (\emph{i.e.} not quantum) relativistic wave equations.  That literature has gone untapped for a number of reasons, including a superficially quantum vocabulary.  That literature is particle physics, of which Wigner's taxonomy of relativistic wave equations in terms of mass and spin \cite{WignerLorentz} is  a prominent example.  The terms ``mass'' and ``spin,'' which misleadingly suggest concepts appropriate to quantum particles rather than relativistic waves, exemplify  the vocabulary issue, on which more below.  While space-time physics ought to be quantization-ready in the sense of recognizing that electrons and other fermions exist (though not much like light and gravity, the usual stars of space-time philosophy \cite{PittsSpinor}) and that classical theories are not the last word, this paper's use of particle physics literature will be entirely as a resource for classical relativistic fields and the space-time philosophy thereof.

In the 1910s Gunnar Nordstr\"{o}m proposed a theory of gravity that met the strictures of Special Relativity \cite{NortonNordstrom,RennGenesis3,vonLaueNordstrom,OBergmannScalar} in the sense of  having, at least, Lorentz transformations as well as space- and time-translations as symmetries, and displaying retarded action through a field medium, as opposed to Newtonian instantaneous action at a distance. This use of the 10-parameter Poincar\'{e} symmetry group reflects a Kleinian subtractive strategy of gradually depriving coordinates of physical meaning \emph{via} symmetries, as opposed to a Riemannian additive  strategy \cite{NortonKleinRiemann}. There  is a larger group  of potential symmetries that one might contemplate, namely, the Bateman-Cunningham 15-parameter conformal group \cite{BatemanConformal0,BatemanConformal,CunninghamConformal0};  Nordstr\"{o}m's theory, which is massless spin $0$ in terms of particle physics, is invariant under that group, whereas massive theories are strictly Poincar\'{e}-invariant.  Nordstr\"{o}m's scalar gravity was a serious competitor to Einstein's program for some years during the middle 1910s.  Neglecting time dependence and nonlinearity, it gives Poisson's equation just as Newton's theory does.  Nordstr\"{o}m's  theory was eclipsed first by the theoretical brilliance of Einstein's much more daring project and the latter's better treatment of Mercury in 1915 (though a ``dark matter'' patch might have been possible), and then by the empirical success of Einstein's theory in the bending of light in 1919, a result manifestly inconsistent with Nordstr\"{o}m's theory.

It is well known that Nordstr\"{o}m's theory does not bend light \cite{Kraichnan}.  That is an immediate consequence of the conformal flatness of the metric in Nordstr\"{o}m's theory  in geometrical form \cite{EinsteinFokker}. 
 and the conformal invariance of Maxwell's electromagnetism \cite{Wald}: space-time is flat in Nordstr\"{o}m's theory except for the volume element, but light doesn't see the volume element in Maxwell's theory in $4$ space-time dimensions.

While representing gravity primarily by a scalar field is no longer a viable physical proposal, there is a great deal that can be learned, surprisingly, by filling in a hole left by the premature abandonment of Nordstr\"{o}m's scalar gravity theory due to Einstein's inventing General Relativity (GR) `too soon.'  While it is evident to  particle physicists that Einstein's theory would have arisen eventually without Einstein (see, \emph{e.g}, \cite{Feynman}), Hans Ohanian, author of a General Relativity textbook \cite{Ohanian} and not a particle physicist, has been prepared to offer, along with some vigorous opinions, even a fairly specific date:  \begin{quote} 
\ldots [I]f Einstein had not introduced the mistaken Principle of Equivalence and
approached the theory of general relativity via this twisted path, other physicists would have discovered the theory of general relativity some twenty
years later, via a path originating in relativistic quantum mechanics. \cite[p. 334]{OhanianEinsteinMistakes}. \end{quote}
Personally I can imagine it perhaps taking as long as 30 years, but one mustn't be too particular about dates in counterfactual history.  In any case the task at hand is to learn what could have been learned in the intervening 20-30 years of that counterfactual history \emph{before} Einstein's equations were found.  
Scalar gravity has  the disadvantage of having been empirically falsified in 1919, but that isn't as bad as it sounds---witness the ongoing reflections on scalar gravity by physicists, often with no particular philosophical or historical interests \cite{GurseyNordstrom,LittlewoodNordstromCosmology,WellnerSandri,DowkerScalar,HarveyScalar,PietenpolScalar,ShapiroScalar,MisnerScalar,BaumgarteShapiroScalar,CalogeroScalar,CalogeroNordstromVlasov,SundrumScalar,LiberatiNordstromEmergent,ReuterManrique,DeruelleScalar,GarrettBinaryScalar,DeruelleConformal,RomeroWeylFrame,CalogeroVlasovNordstromFP}.

Thinking seriously about scalar gravity helps one to separate the wheat from the chaff in Einstein's arguments. For example, as early as 1907  Einstein concluded that a relativistic field theory of gravity could not describe gravity with a scalar potential. In the common sympathetic Einstein historiography,
this conclusion is often presented as a result, or at least isn't challenged.  As it happens, Einstein's argument was wrong \cite{GiuliniScalar}:  
\begin{quote}  
On his way to General Relativity, Einstein gave several arguments as to why a special-relativistic
theory of gravity based on a massless scalar field could be ruled out merely on grounds of theoretical
considerations. We re-investigate his two main arguments, which relate to energy conservation and
some form of the principle of the universality of free fall. We find such a theory-based \emph{a priori}
abandonment not to be justified.  Rather, the theory seems formally perfectly viable, though in clear
contradiction with (later) experiments.   \cite{GiuliniScalar} 
\end{quote} 
  Einstein here seems to have made a lucky mistake, a habit to which Ohanian calls attention.
\begin{quote}
HOW MUCH OF AN ADVANTAGE did Einstein gain over his colleagues by his mistakes?  Typically, about ten or twenty years. \cite[p. 334, \emph{sic}]{OhanianEinsteinMistakes}
\end{quote}
  There would have been much less reason to consider a tensor theory of gravity so early without erroneous arguments against scalar gravity.

Giulini illustrates two important themes:  both the \emph{a priori} plausibility of a graviton mass (to borrow quantum terms for a classical context) and the haste in which the idea is typically eliminated on narrowly empirical grounds, as though nothing conceptually interesting lay in the possibility of a small but nonzero graviton mass.  
\begin{quote} In modern terminology, a natural way to proceed would be to consider fields according to
mass and spin, [footnote suppressed] discuss their possible equations, the inner consistency of the mathematical
schemes so obtained, and finally their experimental consequences. Since gravity is a
classical, macroscopically observable, and long-ranged field, one usually assumes right at
the beginning the spin to be integral and the mass parameter to be zero. The first thing to
consider would therefore be a massless scalar field. What goes wrong with such a theory?
\cite{GiuliniScalar} \end{quote}
That route leads first to Nordstr\"{o}m's theory (massive spin-$0$), but there is no compelling reason to think that the mass is $0$ rather than just small. 
In entertaining scientific theories, one might wish to consider not only trying to get to the truth (in terms of prior probability and evidence), but also utilities \cite{MaherBetting}.  While there are other kinds of utilities,  the utility of being philosophically interesting is especially relevant in this context.  Grinding out one more decimal place might or might not be very important, depending on the details.  Keeping a \emph{precise} count of the number of pigeons in Chicago, though it would yield ecological facts, seems not worth the effort. But adding a mass term in gravity has a high philosophical utility in that one can thereby make a large conceptual-metaphysical difference with an arbitrarily small empirical difference  \cite{UnderdeterminationPhoton}.  Such theories can show us that what seemed to be inevitable philosophical lessons given current scientific knowledge, might in fact be optional.  If today's lessons are not optional because other paths would have converged on them, the lessons might still be justified by reasons other than those usually given.  

This introductory section will close with an outline of what is to come.  One section gives further background on relativistic wave equations and particle physics, including potentially confusing but standard and nearly unavoidable quantum terminology for theories that might be classical.  If (following Pascual Jordan) one expects to take a classical field theory and quantize it, it doesn't seem important to avoid quantum words that apply literally to the expected finished product and that have clear classical analogs.  Subsequent sections consider various major issues in space-time philosophy and show how massive scalar gravity, either in itself or in the strong hints that it gives for massive spin-$2$ gravity (which relates to General Relativity as massive scalar gravity relates to Nordstr\"{o}m's theory) or both, gives a very different perspective.  Gravity could have fit within Minkowski space-time (with only the Poincar\'{e} symmetry group) if gravity had been a massive scalar field.  The apparent explanatory utility of geometrical descriptions of gravity depends mostly upon considering only a narrow collection of theories with the special property of having only one of each type of geometric object plausibly related to geometry (metric, volume element, connection, projective connection, \emph{etc.}); in general it is much more helpful to consider the laws or (more or less the same thing assuming that a variational principle exists) the Lagrangian density.  
The mass term is a tool for violating pretty much all of Einstein's famous Principles, at least in the strong senses that imply interesting conclusions.  Mass terms being plausible, Einstein's Principles are correspondingly less plausible than one might have thought (even given their supposed empirical success).  Debates on conventionalism (including critiques of Poincar\'{e} by Eddington and of Gr\"{u}nbaum by Putnam \emph{et al.}) are seen to depend crucially on what one takes the relevant modal scope of the discussion to be.  Conventionalism at its best considers a broad modal scope including multi-geometry theories, and invokes universal forces only in the less friendly special case of single-geometry theories, whereas the opposing views appear simply to ignore the possibility of multi-geometry theories. An appendix reconsiders the Malament-Weatherall-Manchak view that conventionality of the space-time metric must be restricted to conformally related metrics, leaving only the volume element as conventional.  Taking Thomas's conformal-volume decomposition of a metric into its irreducible parts and applying it to \emph{both} metrics in Reichenbach's conventionality formula $ g'_{\mu\nu} +  F_{\mu\nu} = g_{\mu\nu}$, one can construct the `universal force' that relates the two conformal structures, evading the claimed conformal restriction. 


\section{Wigner's Mass-Spin Taxonomy of Relativistic Wave Equations} 


The range of options in relativistic classical field theory can be found in work from the 1920s-30s in the guise of relativistic  quantum wave equations:  one merely needs to interpret the ``wave function'' as a classical field.  Indeed an older physics idiom used the term ``second quantization'' to reflect the fact than a wave equation that one perhaps initially met as a quantum particle's wave equation (the Klein-Gordon equation, Dirac equation, or the like), was mathematically just a classical field.  Thus ``second-quantizing'' such a relativistic wave equation would give a quantum field theory, while leaving it alone would give a classical field theory.  Nowadays, with the triumph of Jordan's subsumption of all matter into fields, one rarely hears of anything being ``second-quantized.'' But works on quantum field theory still contain large warmup exercises in classical field theory---often the first few chapters of a book, with many other relevant bits scattered throughout.  Likewise, one can learn important things about relativistic classical field theory by reading articles nominally about quantum field theory in such paradigmatically particle physics-oriented venues as \emph{Nuclear Physics B}.  Indeed much of what one learns about ``Special Relativity'' from particle physics literature, though perfectly classical in nature, is harder to learn elsewhere, especially in newer literature. Some examples might be the parity-flipping off-brand tensors (pseudo-scalars, axial vectors, \emph{etc.}), the 15-parameter conformal group and its association with masslessness, massive wave equations, spinors, irreducibility of representations, Belinfante-Rosenfeld equivalence of canonical and metric stress-energy tensors, and two newer examples, the ``improved'' energy-momentum tensor \cite{ImprovedEnergy} and nonlinear group realizations \cite{OPspinor,ColemanWZ,IshamSalamStrathdee}. One finds the divide overcome primarily in work on supergravity (including strings!) and on gauge theories of gravity.  Thus neither special nor general relativistic physics is easily fully and accurately comprehensible without attention to particle physics, notwithstanding various traditional-institutional arrangements in physics, philosophy and technical history of science, especially literature consumption habits.

 Indeed one thing that one learns from particle physics literature is that while a theory might be known to be invariant under the 10-parameter Poincar\'{e} symmetry group ($3$ boosts, $3$ rotations, and $4$ translations for the part connected to the identity), additional symmetries sometimes arise indirectly \cite{Feynman,Deser,OgievetskyLNC}.  Such additional symmetries might be as mild as the $15$-parameter Bateman-Cunningham conformal group, or as wild as the gauge symmetry of Einstein's General Relativity.  There is a respectable usage according to which theories invariant under the $15$-parameter conformal group do not fit within  Special Relativity  \cite[p. 179, 187-189]{MTW}  \cite[p. 19]{NortonNordstrom}, because they do not exhibit the full Minkowski space-time structure.  Hence one might initially think that one has a special relativistic theory, and then discover that one doesn't after all.  Such a shift might feel unsettling, or seem impossible, if one is wedded to a Riemannian ``additive'' strategy \cite{NortonKleinRiemann} of starting with a manifold and adding structures (straight paths, parallel transport, length, \emph{etc.}) bit-by-bit, because adding structures and pronouncing them ``real'' sounds so permanent.  Such a shift is, however, quite natural in terms of a Kleinian ``subtractive'' strategy in which progressively larger symmetry groups strip away reality from structures antecedently considered meaningful.  On the other hand, Kleinian subtractions can feel permanent as well, as in the elimination of Lorentz's aether on the way to Einstein's version of Special Relativity and the elimination of preferred coordinate systems on the way to General Relativity.\footnote{Whether preferred coordinate systems were fully eliminated is not entirely obvious in the wake of the failure of the Anderson-Friedman absolute objects program to yield the expected conclusion that General Relativity is substantively generally covariant \cite{FriedmanJones,GiuliniAbsolute}, the culprit being $\sqrt{-g}.$ }

 One lesson here is that it isn't the case that the Kleinian picture is obsolete; but neither should one embrace Klein and reject Riemann.  Rather than thinking of either Riemannian addition or Kleinian elimination as a strategy for making irreversible progress, it is helpful to think in both directions without prejudice, as moves that can be made depending on the circumstances.  One should be prepared to run the Riemannian additive strategy \emph{in reverse} if the need arises, if a formerly ``real'' structure proves superfluous due to a new Kleinian argument.  To that end, it can be helpful to avoid such reifying terms as ``Special Relativity'' and ``Minkowski space-time,'' which, not coincidentally, are nouns,\footnote{Worries about reification, misplaced concreteness, hypostatization, and the like are not new, but they find a nice example in space-time philosophy.} in favor of adjectives such as ``Poincar\'{e}-invariant'' or perhaps ``special relativistic.'' A useful corrective is thus found in such a title as ``Minkowski Space-time:  A Glorious Non-entity'' \cite{BrownPooleyNonentity}.  But evidence could come to support a new Riemannian addition undoing a Kleinian subtraction.  Such would occur if a satisfactory massive graviton theory were devised and then empirically confirmed (or it could occur in some other way).  Scalar gravity being obsolete, that could only happen for massive spin-$2$ gravity, which has seen renewed attention in the last 15 years (after a dearth from 1972) and an explosion of work since 2010 \cite{HinterbichlerRMP,deRhamLRR}, but it brings up many subtleties as well.  
Indeed physicists have recently contemplated theories in which not only do two metrics exist, but both couple to matter \cite{SolomonBigravityFinsler}; in general there exists no effective pseudo-Riemannian metric describing what  exists, but one can define an effective Finslerian metric, for which the infinitesimal Pythagorean theorem involves a quartic form and hence a symmetric rank-$4$ metric tensor. My task at hand is much simpler because of the assumption that matter sees only one metric (as usual in massive spin-$2$ gravity), and for spin-$0$ gravity the two metrics are conformally related.

Additional insight from the particle physics side of classical relativistic field theory pertains to two modifications that one might envisage making to the wave equation $$\partial_{\mu} \partial^{\mu} \phi = (-\partial^2/\partial t^2 + \nabla^2)\phi=0$$ for a source-free wave equation satisfied by nearly any physical potential $\phi$ (which might be a scalar, a vector, a spinor, \emph{etc.}).\footnote{One can of course also add sources---charge density, energy-momentum density, or the like---to the right side of the equation.  It might turn out that (as in Brans-Dicke gravity, \emph{e.g.}) it isn't terribly clear whether some terms belong on the right side as sources, or on the left side akin to $\partial_{\mu} \partial^{\mu} \phi.$  The choice might be merely conventional, especially classically \cite{FaraoniNadeau}.  Arguably (though less convincingly), there is no fact of the matter in General Relativity either, which absence can be useful  \cite{Kraichnan,Deser,SliBimGRG,DeserRedux}. More generally, one can include interactions and even self-interactions, making the equation nonlinear.}
In the static, spherically symmetric case in spherical polar coordinates, one has
$$  \frac{1}{r^2} \frac{\partial }{\partial r}\left(r^2 \frac{\partial \phi}{\partial r} \right)=0$$ outside sources, giving (with reasonable boundary conditions) a $\frac{1}{r}$ potential and, after taking the derivative, a $\frac{1}{r^2}$ force.  This is familiar, but worth saying for comparison to two alternatives, especially because the potentials are more tractable than the forces.

A modification that tends to go unconceived in the context of General Relativity, but is routine  in particle  physics, involves adding an \emph{algebraic} term in $\phi$ in the field equations.
 The coefficient of such an algebraic term, if the sign is suitable, is called the ``mass'' (squared) of the particle/field $\phi.$  Such terminology makes inessential  use of Planck's constant to achieve proper units; instead one could simply regard the quantity as a new \emph{inverse length scale}, something that many particles/fields demonstrably have (weak bosons, electrons, nowadays at least some neutrinos, \emph{etc.}, less fundamental entities such as protons and neutrons, and various less famous particles, whether fundamental or composite), and by analogy, presumably might be had by any particle/field. 
The resulting wave equation, which was invented multiple times around 1926 \cite{KraghKleinGordonManyFathers}, is known as the Klein-Gordon equation 
$$  (-\partial^2/\partial t^2 + \nabla^2-m^2)\phi=0.$$ 
``Particle mass'' is just a property of a classical field, expressed in entrenched quantum terminology, for which there is no brief alternative.
 In the static, spherically symmetric case, this equation becomes
$$ (\nabla^2-m^2)\phi= \frac{1}{r^2} \frac{\partial }{\partial r}\left(r^2 \frac{\partial \phi}{\partial r} \right) -m^2 \phi =0.$$ For a massive theory, one gets a faster exponential fall-off as $\frac{1}{r}e^{-mr}$. More specifically, a graviton mass $m$ gives  the potential
$-\frac{G M}{r} e^{-m r},$ which gives a attractive force  that is qualitatively similar to the more common $m=0$ case, proportional to the heavy body's mass $M$ and merely weakening faster with distance thanks to the graviton mass $m$. 
 If one pays attention to units---a good habit enforced upon beginning students and later lost by theorists---one notices that $m$ functions as an inverse length.  If one takes $m$ to be really a mass, then $\frac{m c}{\hbar}$ is an inverse length; setting $c=1$ and $\hbar=1$ as usual makes mass and inverse length the same; the length is the reduced Compton wavelength.  A classical theory should know nothing of $\hbar,$ however, so one can take $\frac{m c}{\hbar}$ as primitive, a new inverse length in the wave equation.  Using units such that  $c=1$ and $\hbar=1$ will remove the need to make such distinctions.  
A  $\frac{1}{r}e^{-mr}$ potential appeared in or before the 1890s in astronomy and physics in the works of Seeliger and Neumann\footnote{Norton has insightfully discussed the problem of Neumann's 1890s priority claim with a brief reference to Neumann's 1870s work \cite{NortonWoes}, the problem being that no such Neumann work seems to exist.  In particular (\cite{Neumann1874}) does not seem to be the right paper. I note that Neumann's 1886 paper in a sister journal \cite{Neumann1886} at least is relevant and appeared some years before the mid-1890s, though it isn't from quite the right journal, the right decade, or the right pages.} 
  \cite{Neumann1886,PockelsHelmholtzEquation,Neumann,Seeliger1896,NortonWoes} and again due to Yukawa in particle physics in the 1930s \cite{Yukawa}.  The inverse of $m$ is known as the range of the field, so nonzero $m$ gives a  field a finite-range, while $m=0$ gives a ``long'' or ``infinite'' range.  For the electro-weak theory, for example, the weak nuclear force is not noticeable in daily life as electromagnetism is because the weak force is massive and hence short-ranged, though its mass arises in a more subtle way attributed to the Higgs particle/field.  While simply adding a mass term works fine for electromagnetism even under quantization, and works classically for Yang-Mills, there are distinctively quantum field theoretic reasons (trouble at 1-loop \cite{Slavnov,DeserMass,HurthMassiveYMNoHiggsNonunitary}) for introducing the Higgs particle in Yang-Mills.  Even so, the Higgs particle gives rise, after a field redefinition suited to the true minimum energy, to an effective mass term for the vector bosons.  Now that the Higgs is empirically confirmed, it is still worthwhile to recall why it `had to be there' before it was seen.

A graviton mass term violates the supposedly fairly generic template $$OP(POT)=SOURCE$$ \cite{Renn,RennSauerPathways} ($SOURCE$ being some kind of source term involving mass-energy, $POT$ being the gravitational potential, and $OP$ being a purely second-order differential operator) that Einstein employed in searching for his field equations.  If one comes to believe that matter is relativistic fields, one will want a mass term (inverse length scale) in the wave equation in order to have matter that can sit still, like a tree, rather than travel at the speed of light.  (We don't, of course, presently observe light or gravity sitting still in that sense; perhaps they can't, but that isn't yet clear.)  While the concept of adding a `particle mass' term in the modern sense was not fully available in the 1910s, the analogous concept was entertained for relativistic photons (which lacked the Seeliger-Neumann precedent) already in the early 1920s  by de Broglie \cite{deBroglieBlack,deBroglieWavesQuanta,deBrogliePhilMag}.  During the 1920s-30s progress in relativistic quantum theory, the concept of adding a mass term to the wave equation would become routine.  
  Massive photons were explored initially Proca in terms of fields in 1930s, and later Schr\"{o}dinger and others \cite{Proca,deBrogliePhoton1,deBrogliePhoton2,SchrodingerPhoton1,SchrodingerPhoton2,BelinfanteProca,BassSchroedinger,ProcaHistory}.

  In massive electromagnetism,  the kinetic term $-\frac{1}{4} F_{\mu\nu} F^{\mu\nu}$ in the Lagrangian density has a gauge symmetry, but the  mass term  $-\frac{1}{2} m^2 A^{\mu}A_{\mu}$ breaks the gauge symmetry.  Massive theories also have the advantage of being local in terms of the true degrees of freedom, unlike gauge theories \cite{Sundermeyer}, perhaps giving the best of both worlds to some degree.    
 In terms of quantization, broken gauge theories are somewhat special in relation to naturalness, inheriting some of the benefits of the symmetry that they almost  have \cite{tHooftNaturalness,NaturalnessUnderStress}. 
Strikingly, 't Hooft exempted gravity from naturalness because the cosmological constant was already known to violate it.  On the other hand, a small spin-$2$ graviton mass should be  compatible with naturalness because the massless theory (General Relativity) is more symmetric than the massive theory(s).  While typical scalar field theories do not become more symmetric with $0$ mass, massive Nordstr\"{o}m scalar gravities \emph{do} become more symmetric, shifting from the Poincar\'{e} group to the 15-parameter conformal group.  
't Hooft notes that for scalar theories ``[c]onformal symmetry is violated at the quantum level.''  However, he argues that one have a self-interacting massive $\phi^4$ theory that is natural as long as  the self-interaction is small (because in a free theory particles are conserved) or the mass and self-interaction are \emph{both} small.  What seems to be  excluded is for the mass to be small but the self-interaction large. Fortunately the self-interaction terms for universally coupled massive scalar gravities are of the form $m^2 \sqrt{G}^{j-2} \phi^j$ ($j\geq 3$) \cite{OP,PittsScalar}, implying that the self-interaction is also small if the graviton mass is small.   
 The massless limit of massive electromagnetism (sometimes called a neutral vector meson if the electromagnetic interpretation is not emphasized) is smooth not only in classical field theory \cite{Jackson}, but also in quantum field theory \cite{BelinfanteProca,Glauber,BassSchroedinger,StueckelbergMasslessLimit,BoulwareGilbert,BoulwareYM,GoldhaberNieto,SlavnovFaddeev,DeserMass,GoldhaberNieto2009,Shizuya1,Ruegg,Slavnov},
yielding an interesting case of the underdetermination of theories by data with a non-standard logical form \cite{UnderdeterminationPhoton}.


 Inspired by de Broglie and Pauli-Fierz, Marie-Antoinette Tonnelat and G\'{e}rard Petiau explored massive gravitons in the 1940s  \cite{TonnelatFlux,TonnelatSecond,TonnelatFusion,TonnelatWaves,TonnelatInterpretation,PetiauCR41a,PetiauCR41b,PetiauCR41c,TonnelatSecondRank,Tonnelat,deBroglie1,TonnelatDisquisitiones,PetiauCR43a,PetiauCR43b,PetiauRevue,Tonnelat20,TonnelatGravitation,TonnelatNewtonian,PetiauCR44a,PetiauCR44b,PetiauRadium45,PetiauRadium46a,PetiauRadium46b,deBroglie}. The gravitational case is in fact earlier, due to Neumann, Seeliger and Einstein.  
  Massive theories are plausible in terms of relativistic field theory.  
 As Freund, Maheshwari and Schonberg put it, 
\begin{quote}
In the Newtonian limit, equation (1) is now replaced by the Neumann-Yukawa
equation, 
$$ (\Delta -m^2)V = \kappa \rho \hspace{1.2in}  (3),$$
which leads to the quantum-mechanically reasonable Yukawa potential 
$$ V(r) = - \frac{ \kappa M e^{-m r} }{r}, \hspace{1in} (4) $$
rather than the peculiar oscillator [due to the cosmological constant $\Lambda$] of  equation (2).  \cite{FMS}. 
\end{quote} 
This potential was sufficiently plausible as to be independently invented 3 times (Seeliger among many other potentials, Neumann, and Einstein); Seeliger and Einstein were both addressing the problem of mathematically divergent gravitational potential in an infinite homogeneous static Newtonian universe.  
The peculiarity of $\Lambda$ and its resistance to sensible interpretations has also been noticed by authors who do not contrast it with a graviton mass and who do not see its peculiarities as  \emph{reductios}  \cite{McCreaLambda,KerszbergInvented}.
For massive gravitons one has the plausible form $$OP(POT)+POT=SOURCE,$$ which is excluded by the narrow schematic equation employed by Einstein with no $POT$ term permitted.

In the first half of his 1917 paper on the cosmological constant $\Lambda,$ Einstein briefly entertained what is in effect a massive scalar gravitational theory: 
\begin{quote}
We may ask ourselves the question whether [these difficulties involving the Newtonian potential in a cosmological context] can be removed by a modification of the Newtonian theory.  First of all we will indicate a method which does not in itself claim to be taken seriously ;  it merely serves 
as a foil for what is to follow.  
In place of Poisson's equation we write
\begin{eqnarray*} 
\hspace{2cm} \nabla^2 \phi - \lambda \phi = 4 \pi \kappa \rho  \hspace{1.25cm}  .  \hspace{1.25cm}  .  \hspace{1.25cm}  .  \hspace{1.1cm} (2)
\end{eqnarray*}
where $\lambda$ denotes a universal constant.  \cite[p. 179]{EinsteinCosmological}
\end{quote}  
 Thus Einstein in effect contemplated a theory of the sort that, in light of later quantum mechanics terminology, one might call a theory of gravity using a nonrelativistic massive scalar field \cite{DeserMass}, with $\lambda$ equaling the square of the scalar graviton mass (with $\hbar = c=1$).  Relativistic massive scalar fields in the absence of interacting satisfy the Klein-Gordon equation, but interpreting the field as gravity introduces interactions, including self-interaction and hence nonlinearity.  %

Einstein's cosmological constant $\Lambda$ has waxed and waned in its empirical fortunes, but its plausibility or implausibility \emph{vis-a-vis} relativistic wave equations has not always been appreciated.    Unfortunately, the dominant effect of $\Lambda$  is to introduce a \emph{constant} into the field equations, like  $  \frac{1}{r^2} \frac{\partial }{\partial r}\left(r^2 \frac{\partial \phi}{\partial r} \right) + C=0$.
Widespread historical misunderstanding of this fact, going back to Einstein in 1917 \cite{EinsteinCosmological}, has occurred.  Such an alteration seems likely to produce a rather peculiar point mass potential.  Indeed it does:  the potential grows quadratically with distance, like a harmonic oscillator (or its opposite, depending on the sign), and not at all like any fundamental physical force behaves in mundane experience.  (The strong nuclear force is not part of what I mean by mundane experience; intrinsically nonperturbative theories are different.) The matter was well described by  Freund, Maheshwari and Schonberg. 
\begin{quote}
In the ``Newtonian'' limit it leads to the potential equation,
$$\Delta V + \Lambda =  \kappa \rho.  \hspace{1.3in}  (1)$$
Correspondingly, the gravitational potential of a material point of mass $M$ will be given by
$$ V   = -\frac{1}{2} \Lambda r^2 - \frac{\kappa M}{r}. \hspace{1in}  (2) $$
A ``universal harmonic oscillator'' is, so to speak, superposed on the Newton law. The
origin of this extra ``oscillator'' term is, to say the least, very hard to understand.
\cite{FMS}  \end{quote}
Such a modification, like a graviton mass term, also violates the supposedly fairly generic template $$OP(POT)=SOURCE$$ \cite{Renn,RennSauerPathways} that Einstein employed in searching for his field equations, because  a cosmological constant involves the gravitational potentials algebraically. For $\Lambda$ one has 
the curious form  $$OP(POT)+POT+CONST=SOURCE,$$ because the gravitational potential/field indicating deviation from triviality is something like $g_{\mu\nu}-diag(-1,1,1,1).$ 
Both terms $POT$ and $CONST$ are novel, but $CONST$ is abnormal.

Tragically, Einstein conflated these two quite different ideas, $\Lambda$ and (what we now call) a graviton mass, in his famous 1917 paper that introduced $\Lambda$ \cite{EinsteinCosmological}. 
Many, many authors have added a cosmological constant to General Relativity and thought, erroneously, that they had thereby given the graviton a mass.  But it makes all the difference whether the lowest order algebraic term, which will dominate for weak fields, is zeroth order within the field equations (first order within the Lagrangian density), as with the cosmological constant, or first order within the field equations (second order within the Lagrangian density), as with a mass term.  Not only the gross qualitative behavior of the solutions, but also the presence or absence of gauge freedom, are at issue:  mass terms tend to remove gauge freedom in favor of having more physical degrees of freedom  \cite{PauliFierz,Fierz2}.  The cosmological constant $\Lambda$ does not remove gauge (coordinate) freedom and hence does not reduce the number of physical degrees of freedom.   Hence neglecting the massive possibility can lead to overconfidence in the existence of a large symmetry group. 
 By conflating the cosmological constant $\Lambda$ in General Relativity with (what we now construe as) a graviton mass, Einstein helped to obscure for himself and others the deep conceptual issues raised by the mass term.

A substantial portion of 20th century physics was not seriously attended by philosophers of space-time either at the time or later.  According to Wes Salmon, ``[d]uring the years between 1930 and 1950, roughly, little of significance seems to have been achieved in philosophy of space and time.''  \cite[p. 29]{SalmonReichenbachLogicalEmpiricist}. While this assessment seems true, it is no reflection on what physicists were doing.  Rather, it says more about how only a few philosophers were productively involved in studying physics in the 1920s, and then they stopped. Schlick and Carnap had turned their attention elsewhere well before the end of the decade (followed by Schlick's assassination in the 1930s).  Even Reichenbach, who paid more serious technical attention to physics for longer \cite{Reichenbach1929Zeitschrift}  than one might expect from reading the truncated English translation of his book \cite{ReichenbachSpace}, quit paying much attention after \emph{c.} 1930.  So crucial a development as the inclusion of fermions (including electrons and, less fundamentally, protons and neutrons)---which Weyl took to be a great conceptual novelty due to its conclusion of the inadequacy of tensor calculus \cite{WeylGravitationElectron,ScholzWeylFockSpinor}---fell into the period of neglect.\footnote{The point that philosophers did not keep up with developments in physics is strengthened, not weakened, by the fact that after Salmon's period of philosophical stasis, further physical innovations in the 1960s involving nonlinear group realizations, still poorly known even among physicists outside the supergravity community, largely deflated Weyl's result \cite{OPspinor,IshamSalamStrathdee,BilyalovSpinors,PittsSpinor}.} 
  Given the degree to which space-time philosophy took lasting shape in the late 1910s-20s in the thought and works of Schlick, Reichenbach and Carnap, the fact that Einstein's false analogy went undetected until 1942 in Germany \cite{Heckmann}, and was not challenged again until the 1960s, enabled philosophers to quit paying attention to  physics relevant to space-time (but not necessarily primarily \emph{about} space-time!) long before the issues were sorted out. 
 This oversight has never been corrected, not least due to the general relativity \emph{vs.} particle physics split within physics (on which see (\cite{Feynman,Rovelli})), a barrier across which little communication occurs, except through supergravity \cite{BrinkDeserSupergravity} (and superstrings!) and gauge theories of gravity.  Most philosophers and historians take most of their guidance on space-time from general relativists, so whatever is best learned from particle physicists is less familiar.

In the actual contingent history, Einstein  was unaware of Seeliger's work until  after the final GR field equations were known  \cite[p. 420]{EinsteinSpecial}  \cite[p. 557]{EinsteinForster}  \cite[pp. 142, 146]{EinsteinLunar} \cite[p. 189]{EinsteinNotes}. (\emph{Pace} Earman, \cite{EarmanLambda}, section 30 addressing Seeliger  in Einstein's popular book first appeared in 1918  \cite[p. 420]{EinsteinSpecial}).  When he did discuss the idea in 1917 (not yet aware of Seeliger's work) \cite{EinsteinCosmological}, he drew an  analogy between (what we would call) massive scalar gravity and his cosmological constant $\Lambda$ term, but a spurious one 
 \cite{Heckmann,Trautman,DeWittDToGaF,FMS,Treder,Schucking,NortonWoes,CooperstockTerm,HarveySchucking,EarmanLambda}---an error that would resurface often. 
This false analogy---Einstein's `other' blunder with the cosmological constant (besides the reportedly self-diagnosed blunder of introducing it in the first place \cite[p. 44]{GamowEinsteinLambda})---tends to obscure the possibility of a genuine spin 2 analog to massive scalar gravity.  In 1913 Einstein even enunciated a principle to the effect that the field equations for gravity should not depend on the absolute value of the gravitational potential(s) 
\cite[p. 72]{NortonNordstrom} \cite[pp. 544, 545]{EinsteinPresent}.  It follows immediately that a mass term is not permitted, but there is little justification for the principle.   Modern historians of GR, in the course of commenting on Einstein's principle of simplicity \cite[pp. 501-503]{EinsteinPapers4}, seem unaware of the fact that Einstein in 1913 thereby excluded both massive scalar gravity and massive GR from the list of theories that he would entertain. As Norton notes, Einstein's refusal to take the modified Poisson equation  seriously \cite{EinsteinCosmological} is not accompanied by  good reasons  \cite{NortonWoes}.  Much of his motivation is his \emph{a priori} opposition to absolute inertial coordinate systems \cite{NortonTriumph,NortonFateful}, an opposition that one can  fail to share.  Even if one shares it, one loses the possibility of supporting this opinion by evidence if one refuses to entertain and critique theories that contradict it.

 Part of the novelty of the treatment below consists in pointing out how this conflation lead to total failure until now to recognize the \emph{philosophical} interest of massive gravity.  One could consider either massive scalar gravity  (a cousin to Nordstr\"{o}m's 1914 theory) or massive tensor gravity (a cousin to Einstein's theory).  It turns out that there is more than one massive scalar gravity theory \cite{PittsScalar} and, presumably, more than one massive tensor gravity \cite{OP,MassiveGravity2,HassanRosenNonlinear}, due to the many possibilities for self-interaction (algebraic nonlinear terms),  whittled down to give theories that don't have certain subtle problems.  Massive scalar gravity, which approximates Nordstr\"{o}m's theory arbitrarily well for sufficiently small graviton mass, was thus falsified in 1919 by the bending of light, whether anyone had conceived of it or not.  It was born refuted. Perhaps that is not a rare problem; Lakatos claimed that it is a ``historical fact that most important theories are born refuted'' \cite[p. 114]{LakatosHistory}. Massive scalar gravity will not become an important theory, but its cousin massive spin-$2$ gravity perhaps might.  Thus since 1919 the problem of unconceived alternatives applies more properly in terms of massive spin-$2$ gravity.   Massive scalar gravity is easier to understand (a pedagogical virtue for philosophical consumption), indeed much easier to analyze for working physicists, whose views on the viability of massive tensor gravity shifted radically for the worse in the early 1970s, gradually improved since 1999, and shifted radically for the better in 2010, with continuing change since then.  Whether or not massive tensor gravity ultimately makes sense (an outcome which is difficult to judge in  2015 due to the rapid pace of physical development), massive scalar gravity clearly does make sense.  It has  plenty of lessons for what space-time could have been like, as well as interesting, currently plausible suggestions for what space-time in fact might be like---to the degree that massive tensor gravity works and insofar as massive tensor gravity is analogous to massive scalar gravity (a presumptive analogy that can fail in surprising ways requiring the reinstallation of gauge freedom \cite{MassiveGravity1}!).  Hence right now there is a great deal to learn about space-time philosophy from massive scalar gravity.  The field seems to be entirely open.  The only near-exceptions that come to mind displaying philosophical awareness of massive gravity (apart from (\cite{PittsScalar,UnderdeterminationPhoton})) is  some 1970s work by Peter Mittelstaedt \cite{MittelstaedtLorentz}.  But even Mittelstaedt merely described the graviton mass merely as an empirical parameter that might be $0$, not as a conceptual watershed as it should be seen, and as some physicists recognized \cite{FMS}.  Furthermore, massive gravitons shortly ran into serious trouble \cite{vDVmass1,vDVmass2,DeserMass}.  Recently the tide has turned and massive graviton theories have become a `small industry' \cite{deRhamGabadadze,HassanRosen,HassanRosenNonlinear,HinterbichlerRMP,deRhamLRR}.  Whether or not massive spin-$2$ gravity survives as a viable theory and rival to General Relativity, progress will have been made by exploring serious alternatives.

Particle physics would enable historians of General Relativity to ask  questions that they tend not to ask, such as why Einstein did not seriously consider massive gravities (despite eventually (p)reinventing massive scalar gravity in 1917, in a sense).
 Developments in group theory as applied to relativistic quantum mechanics, such as by Wigner  \cite{WignerLorentz,BargmannWigner},   
 classified all possible fields in terms of the Lorentz group with various masses and various spins. (Merely the words that are quantum mechanical; the concepts are just classical field theory.  Avoiding the quantum words involves using \emph{c.} 5 times as many syllables.)
 As noted above, relativistic massive scalar fields, if non-interacting, satisfy the Klein-Gordon equation
$ (-\partial^2_t + \nabla^2 - m^2) \phi=0, $
where the speed of light and Planck's constant have been set to $1.$
Massive fields with spins higher than $0$ also tend to satisfy the Klein-Gordon equation as a consequence of logically stronger equations of motion.  
 Given particle physicists' taxonomy in terms of mass and spin, it is natural to look for and to fill in the blanks by considering all the possibilities, as the table suggests. The table omits half-integral spins, which have to be fermions by the spin-statistics theorem, and hence do not accumulate into powerful classical forces.  It also omits higher spins, which cannot produce long-range forces due to the lack of suitable conserved currents to which they could couple \cite[p. 253]{WeinbergQFT1}.  

\begin{center}
Mass-Spin Taxonomy Exemplified\\
\begin{tabular}{| c | c | c | c |} \hline
    & Spin 0 & Spin 1 & Spin 2 \\ \hline
$m=0$ & Nordstr\"{o}m & Maxwell & Einstein \\ \hline
$m \neq 0$ &  ?   & de Broglie-Proca        &    ?   \\  \hline 

\end{tabular} 

\end{center}


 In the late 1930s Pauli and Fierz found that the theory of a non-interacting massless spin 2 (symmetric tensor) field in Minkowski space-time was just the linear approximation of Einstein's GR \cite{PauliFierz,FierzPauli,Wentzel}.
 Tonnelat and Petiau, associated with de Broglie pursued massive spin 2 theories (cited above). Tonnelat cited Fierz \cite{TonnelatWaves,Fierz}. Thus by the end of the 1930s,  the idea of a graviton mass was  available not merely by analogy to electromagnetism, or the older non-relativistic work by Neumann, Seeliger, and Einstein, but in detailed relativistic work in several papers by a leading physicist (Pauli) with gravity as one intended application, with follow-on work in France encouraged by de Broglie in the early 1940s.

Nordstr\"{o}m's theory of a long-range scalar field is, in this particle physics terminology applied retrospectively, a theory of a massless spin $0$ field; thus when one considers  Nordstr\"{o}m's theory, it is natural to consider a massive variant and to ascertain whether the massless limit of the massive theory is smooth.  If it is, then the massive variant serves as a rival to the massless theory, implying a case of underdetermination of theories by data.  As Boulware and Deser put it, there is a 
\begin{quote}
basic principle, physical continuity, which demands that a theory be ``stable'' in its predictions, i.e., no more isolated
from nearby models than our finite observations warrant. In particular, a good theory of long-range forces should have a smooth limit as 
the range tends to infinity, and this limit should agree with the strictly infinite-range model. This
viewpoint has been forcefully stated for electrodynamics
by Schr\"{o}dinger,\footnote{Reference to \cite{BassSchroedinger}. }  and it has been amply
demonstrated by analysis of massive vector theory [references suppressed]  that approximate gauge invariance is not the
contradiction it first seems.
\cite{DeserMass} \end{quote}
(This was not their final view in light of the new Yang-Mills and gravity cases, but it sets up the appropriate expectation in terms of particle physics knowledge through the 1960s, which is progress.)  
Massive scalar gravities, if the mass is sufficiently small, fit the data as well as does Nordstr\"{o}m's theory, as a consequence of the smoothness of the limit of a massive scalar field theory as the mass goes to zero  \cite[p. 246]{WeinbergQFT1} \cite{DeserMass}.  
 Thus there is a problem  of underdetermination between the massless theory and its massive variants for sufficiently small masses \cite{UnderdeterminationPhoton}.

This instance of underdetermination, apart from framing in terms of particle mass,  was already clearly anticipated by Seeliger in the 1890s. He wrote  (as translated by John Norton) that Newton's law was ``a purely empirical formula and assuming its exactness would be a new hypothesis supported by nothing.'' \cite{Seeliger1895a,NortonWoes}
That claim is too strong, in that Newton's law had virtues that not every rival formula empirically viable in the 1890s had.  But a certain kind of exponentially decaying formula was associated with an appropriate differential equation  and hence had theoretical credentials comparable to Newton's \cite{PockelsHelmholtzEquation,Neumann}, vindicating the spirit of Seeliger's point.   
The idea of exploring whether a massive theory could work in place of a massless one (or \emph{vice versa}), much as Seeliger proposed, is a commonplace in particle physics.

 The massless \emph{vs.} massive competition is an especially interesting and well motivated example of the fact, noted by Pierre Duhem, that the curve fitting problem always applies in physics: through any set of experimental results (especially with error bars!), multiple curves can be proposed as the correct theory.  Two consecutive section headings from the famous  part II, chapter 5 of Duhem's book make the point:  ``A Law of Physics Is, Properly Speaking, neither True nor False but Approximate'' and ``Every Law of Physics Is Provisional and Relative because It Is Approximate'' \cite[pp. 168, 172]{Duhem}. There are many ways that a given body of data can be fit by a theoretical formula, but Duhem expects that generally a choice of one option will be made on the basis of good sense.  However, the competition between massive and massless theories is one that good sense does not settle, in that both competitors are taken seriously by particle physicists until reason to the contrary is found \cite{DeserMass}.  While successes of the ``gauge principle'' since the early 1970s are noteworthy and encourage gauge freedom and masslessness for spins high enough to imply gauge freedom ($\geq 1$), the moderately surprising and fairly recently learned fact that at least some neutrinos are massive \cite{NeutrinoMassPramana,NeutrinoMassLNP} serves as a reminder not to neglect massive theories.  Spin $0$ and spin $\frac{1}{2}$ particles have no negative-energy lower spin degrees of freedom that one might want to get rid of \emph{via} gauge freedom, because one cannot take a divergence or trace to make lower spin.  Thus the gauge principle does not apply.

Logically speaking, mass terms are children not of quantum theory or whatever quantum `particles' might be, but of special relativity, the idea that ponderable matter is made of fields (arising from ideas of Mie, Hilbert and Jordan), and daily experience.  Whether or not one has the particle-related concept of a particle mass, it is empirically obvious that most stuff doesn't move at the speed of light or look like standing waves made from oppositely-directed waves moving at the speed of light, for example.  Thus one has an overwhelming empirical motive to look for classical relativistic wave equations with dispersion and the possibility of having the bulk of matter be motionless in some reference frame.  Thus the introduction of the $\frac{\partial^2}{\partial t^2}$ terms due to relativity does much to motivate the algebraic terms.


\section{Massive Scalar Gravity Is Just Special Relativistic}

Features of Nordstr\"{o}m's scalar gravity  are  said to have shown that even the simplest and most conservative relativistic field theory of gravitation had to burst the bounds of Special Relativity (SR) \cite[pp. 179, 187-189]{MTW}  \cite[p. 19]{NortonNordstrom}. 
 Relativistic gravity couldn't be merely special relativistic, according to these claims.  Nordstr\"{o}m's theory indeed has a merely conformally flat space-time geometry \cite{EinsteinFokker}, and it arguably is the simplest and most conservative option. But how do such claims fare in light of the broader range of possibilities of particle physics (or Neumann-Seeliger-Einstein), especially with a graviton mass term as an option?

The best way to write the conformally flat geometry---well adapted to ontology by attending to Ockham's razor (no gratuitous introduction of volume elements that one doesn't want and conformal transformations to cancel them out)---involves breaking a metric into its irreducible parts.  (For irreducible geometric objects in differential geometry, see (\cite{ZajtzPrimitive,StachelRelations})).  Having become familiar with the irreducible parts, their physical meanings, and how to do tensor calculus with them, one can then take them to be primitive, regarding the metric as derived, if it exists at all.  One can  use one part without the other, swap one part out and replace it with a different one of the same sort, introduce two or more of the same type, \emph{etc.} 

  It is a classical result  that one can define a tensor density that precisely picks out the conformal part of a metric, excising any information about volumes. According to Kentaro Yano in 1939, 
\begin{quote}
\ldots M.  T. Y. Thomas$^{2)}$ a introduit, en 1925, une densit\'{e} tensorielle du poids $-\frac{2}{n}$ 

(0.4) \hspace{1in} $G_{ij}=g_{ij}/g^{\frac{1}{n} }$ 

o\`{u} $g$ est le d\'{e}terminant form\'{e} avec les $g_{ij}.$  \cite[p. 72]{YanoConformal}
\end{quote}   
The reference is to (\cite{ThomasRelative}); the work (\cite{ThomasConformal}) further explored such matters.  
 This tensor density has a dimension-dependent and (for dimension $3$ or greater) fractional density weight (poids). It follows that the determinant of Thomas's  quantity $G_{ij}$ (which I write with a caret as $\hat{g}_{\mu\nu}$, partly by association with the unit vectors of vector calculus) is $1$ for the positive definite case, or $-1$ for space-time, in all coordinate systems.   Densities of arbitrary weight seem to be due to Veblen and Thomas (who called them relative tensors) \cite{VeblenThomasDensityDerivative} and to Weyl \cite{WeylGroupsMZ1} \cite[p. 462]{HawkinsGroupsBook}. 
Densities acquire an extra weight-related term\footnote{It is perhaps unfortunate linguistically that one has to distinguish, besides the ordinary mass of heavy objects and the associated weight that one determines using a bathroom scale, the unrelated idea of a graviton mass and the further unrelated mathematical idea of density weight. } in their Lie and covariant derivatives \cite{VeblenThomasDensityDerivative,Schouten,Anderson}.  
Differential geometry in the modern style has tended to employ Weyl's conformal rescalings rather than Thomas's more direct and economical characterization of conformal geometry, a tendency critiqued (without the historical context) by Calderbank and Pedersen: 
\begin{quote} As counterpoint to the tendency to do conformal geometry in a Riemannian framework, we would like to suggest that a conformal structure is more fundamental than a Riemannian structure by defining the latter in terms of the former. \cite[p. 391]{Calderbank} \end{quote}   Branson comments that Thomas's work was largely forgotten  until the 1990s \cite[p. 180]{Branson}. This forgetting seems to refer to mathematicians (apart from \cite{SchoutenHaantjesConformal,YanoConformal,HaantjesConformalSpinor}), because physicists often remembered  \cite{PeresPolynomial,Anderson,AndersonFinkelstein,UnruhUGR}).

Writing Thomas's equation for a flat metric tensor $\eta_{\mu\nu}$ (not that flatness affects this decomposition), one has
\begin{equation}
\hat{\eta}_{\mu\nu} = \eta_{\mu\nu}  (-\eta)^{-\frac{1}{4}}
\end{equation} 
after specializing to $n=4$ space-time dimensions.  One can invert to express the metric in terms of its irreducible parts, the conformal (angle-related) part 
$\hat{\eta}_{\mu\nu}$ and the (unsigned) volume element $\sqrt{-\eta}:$ 
\begin{equation} \eta_{\mu\nu}= \hat{\eta}_{\mu\nu} \sqrt{-\eta}^\frac{1}{2}. \end{equation}

One now can and should think of $\hat{\eta}_{\mu\nu}$ and $\sqrt{-\eta}$ as independent entities in their own right.  $\hat{\eta}_{\mu\nu}$ describes a conformally flat geometry---perhaps one should say a flat conformal geometry, flat in the same of vanishing Weyl curvature tensor, ``conformal geometry'' to emphasize that it isn't a geometry that defines distances. The 15-parameter conformal group first studied by Bateman and Cunningham \cite{BatemanConformal0,CunninghamConformal0,BatemanConformal} is just the group of generalized Killing vectors for $\hat{\eta}_{\mu\nu}, $ the vectors for which it has $0$ Lie derivative.  $\hat{\eta}_{\mu\nu}$  determines the light cones just as if for a flat metric in SR. 
  Note that there is nothing ``flat'' about  $\sqrt{-\eta}$, because one volume element is like another.  The flatness of 
$ \eta_{\mu\nu}$, insofar as it goes beyond the (conformal) flatness of $\hat{\eta}_{\mu\nu},$ is due to a relation between $\hat{\eta}_{\mu\nu}$ and $\sqrt{-\eta}$ rather than a property of $\sqrt{-\eta}$.  Using a volume element one can define a ``volume connection,'' a term due to (\cite{ColemanKorteFirstSecond}).  The entity is familiar as the trace of the Christoffel symbols, but it is a gradient and hence has $0$ curvature, the `other' trace of the Riemann tensor, which one usually takes to vanish.

One can write Nordstr\"{o}m's theory geometrically, as Einstein and Fokker showed \cite{EinsteinFokker}.  One can write it more economically using a Thomas-style decomposition with no surplus structure: 
\begin{equation} g_{\mu\nu}= \hat{\eta}_{\mu\nu} \sqrt{-g}^\frac{1}{2}. \end{equation} 
$ \sqrt{-g}$ includes the influence of gravity.  One could derive  $\sqrt{-g}$ by combining the gravitational potential with $\sqrt{-\eta}$ \cite{Kraichnan,DeserHalpern,PittsScalar}; for Nordstr\"{o}m's theory, such a derivation is a plausible heuristic but ultimately perhaps not illuminating, in that $\sqrt{-\eta}$ does nothing by itself in the final theory (at least not locally)---like a Poincar\'{e}-Reichenbach universal force.  Because the only non-variational field in the theory's Lagrangian density is  $\hat{\eta}_{\mu\nu} $, the symmetry group of the non-variational fields is the 15-parameter conformal group, a larger group than the usual 10-parameter Poincar\'{e} group of Special Relativity, thus admitting a larger class of preferred  coordinate systems  than do paradigmatic  special relativistic theories.  Thus not all the structure of Minkowski space-time is exhibited, and the effective geometry seen by rods and clocks is curved and only conformally flat  \cite[pp. 179, 187-189]{MTW}  \cite[p. 19]{NortonNordstrom}.

 But  massive variants of Nordstr\"{o}m's theory contain both $\sqrt{-g}$ and $\sqrt{-\eta}$ in the mass term and so  are merely Poincar\'{e}-invariant,  hence strictly special relativistic, as far as symmetries are concerned.  Rods and clocks are distorted by gravity, \emph{and in a way that can be empirically ascertained due to the mass term}.  
The difference in symmetry group of the non-variational (that is, not varied in the principle of least action \cite{GotayIsenberg,FriedmanJones}) objects reflects the difference in non-variational objects present in the Lagrangian density. Because of the graviton mass term, massive scalar gravities have the full flat background metric $\eta_{\mu\nu}$ in the Lagrangian density \cite{PittsScalar}.  Thus the Lagrangian has only the symmetries of $\eta_{\mu\nu}$, $10$ Killing vectors ($4$ translations, $6$ boost-rotations).  By contrast in the (massless) Nordstr\"{o}m theory, only the conformal part of a flat metric, $\hat{\eta}_{\mu\nu}$, is present in the Lagrangian density.   Consequently the symmetries are the conformal Killing vector fields, those such that the Lie derivative of the conformal metric density\footnote{ One often sees the conformal Killing equation in terms of an arbitrary metric in a conformal equivalence class and require that the Lie derivative of such a metric be proportional to that metric, $\pounds_{\xi} \eta_{\mu\nu} \sim  \eta_{\mu\nu}$, as if it weren't known how to isolate the relevant piece.  But Thomas showed how in the 1920s; the arbitrary volume element is no more relevant to conformal Killing vector fields than is the aether  to Special Relativity. }  vanishes \cite{Schouten,Anderson}:
\begin{equation} 
\pounds_{\xi} \hat{\eta}_{\mu\nu} = \xi^{\alpha} \frac{ \partial }{\partial x^{\alpha} } \hat{\eta}_{\mu\nu}   + \hat{\eta}_{\mu\alpha}  \frac{ \partial }{\partial x^{\nu} }  \xi^{\alpha}    + \hat{\eta}_{\alpha\nu}  \frac{ \partial }{\partial x^{\mu} }  \xi^{\alpha} 
  - \frac{1}{2} \hat{\eta}_{\mu\nu}   \frac{ \partial }{\partial x^{\alpha} }     \xi^{\alpha}.
 \end{equation}

Because a scalar graviton mass term breaks the 15-parameter conformal symmetry and leaves only the 10-parameter Poicar\'{e} symmetry, it is therefore false, \emph{pace} (\cite[p. 179, 187-9]{MTW}  \cite[p. 19]{NortonNordstrom}), that relativistic gravitation could not have fit within the confines of Special Relativity as construed to require exhibiting the full Minkowski space-time structure.  While it is true that no phenomena required the mass term, it was epistemically possible that the mere smallness of the mass parameter explained its empirical obscurity, as Seeliger had already proposed in the Newtonian case.  Indeed it is still possible, or rather again possible (after seeming impossible since the early 1970s), that a tensorial analog of this issue exists today. Something similar happened with neutrinos a few years back, one recalls.  

%


\section{Massive  Gravit\emph{ies}, Laws, Geometry, and Explanation}

A  recent debate among philosophers of physics, albeit one with striking similarities to older debates about conventionality, pertains to the explanatory priority of space-time structure \emph{vs.} the field equations satisfied by fields on space-time.  Harvey Brown has recently defended the latter view  \cite{BrownPhysicalRelativity,BrownRods}, known by terms such as constructive relativity or physical relativity.  While the former view has an undeniable economy and provides in some respects a natural classification---for example, Einstein's General Relativity and theories involving higher powers of the Riemann  tensor  clearly have some natural affinities---consideration of massive theories of gravity reveals that space-time structure, construed as a list of geometric object fields defined on a manifold, is either too narrow or unhelpfully vague \cite{FMS}.

If there is more than one metric---and why shouldn't there be, if one's imagination is fueled by mathematics after the 1910s \cite[chapter 8]{LeviCivita}?---much of the phenomenology is  unspecified until the field equations are introduced.  Even with the modest ingredients of a flat space-time metric and a scalar gravitational potential, uncountably infinitely many theories can be written down due to the possibilities for a mass term, at least a one-parameter family  \cite{PittsScalar,UnderdeterminationPhoton} but likely more.
Indeed one can derive the full nonlinear Lagrangian density for a universally coupled massive scalar gravity theory using a relation of the form 
 $$ \tilde{g} =\sqrt{-g}^w = \sqrt{-\eta}^w + 8 w \sqrt{\pi G} \tilde{\gamma},$$ where $\tilde{\gamma}$ is the gravitational potential and $w$ is an arbitrary real number (the case $w=0$ requiring special care).  If one already has the Lagrangian density in hand, then one is more interested in whether one can write it using 
 only a combined quantity such as $\sqrt{-g}$, or whether one needs both $\sqrt{-g}$ and $\sqrt{-\eta}.$  The  kinetic term (with time and space derivatives of the potential) is just that of Nordstr\"{o}m's theory and so can be written in terms of $\sqrt{-g}$ only; thus Nordstr\"{o}m's theory is geometrizable.  The graviton mass term can be written in terms of the effective volume element $\sqrt{-g}$ (with volumes distorted by gravity) and also the undistorted volume element $\sqrt{-\eta}$ of the flat metric.  The fact that $\sqrt{-\eta}$ appears by itself, not merely hidden within $\sqrt{-g}$ clothed by $\gamma,$ \emph{implies most of what is conceptually interesting about massive scalar gravity \emph{vis-a-vis} Nordstr\"{o}m's theory}, blocking the usual criticisms of universal forces as superfluous.  This $1$-parameter family of massive scalar gravities is analogous to the $2$-parameter family of massive spin-$2$ gravities proposed in 1965 as variants of General Relativity \cite{OP}.  The scalar case has the advantages of technical simplicity and immunity to the difficulties that have at times (especially 1972-\emph{c.} 1999 or 2010) afflicted massive spin-$2$ gravity.  
  It isn't very profitable to ask what the gravitational potential is in relation to  $\sqrt{-g}$ and $\sqrt{-\eta}$, because after the linear term, the answer is merely a conventional choice. (For the expression above, the answer depends on $w.$)  In fact that freedom to use a variety of field (re)definitions is a crucial \emph{resource} in the derivations \cite{OP,MassiveGravity1,MassiveGravity2,PittsScalar,MassiveGravity3}.  Linearity is convenient; equivocating about the meaning of the quantity \emph{in which} an expression is linear lets one discuss infinitely many theories at once.  That is much faster than trying to derive an uncountable infinity of theories one at a time, and much more general than deriving only one theory as in (\cite{FreundNambu,FMS}).  The parameter $w$ is the density weight of the gravitational potential under coordinate transformations used in the initial derivation.

For any real $w$ (using  l'H\^{o}pital's rule for $w=0$ or $w=1$), a universally coupled massive  theory is given by  \cite{PittsScalar}
\begin{equation} 
 \mathcal{L} = \mathcal{L}_{Nord} +   \frac{m^2}{64 \pi G} \left[ \frac{  \sqrt{-g} }{w-1}   +   \frac{ \sqrt{-g}^w \sqrt{-\eta}^{1-w} }{w(1-w)}  -  \frac{ \sqrt{-\eta}  }{w} \right]_.  
\end{equation} 
One can express this mass term as a quadratic term in the potential  and, typically, a series of higher powers using the expansion $\sqrt{-g}^w = \sqrt{-\eta}^w + 8 w \sqrt{\pi G} \tilde{\gamma},$ where $\tilde{\gamma}$ is the gravitational potential.  Different values of $w$ thus give different definitions of the gravitational potential (disagreeing at second or higher order).  If one takes the $w$th root and then takes the limit $w \rightarrow 0,$ the limit is
$$ \sqrt{-g} =\sqrt{-\eta} exp(8 \sqrt{\pi G} \gamma) .$$
An exponential change of variables very much like this was already employed by Kraichnan, though without application to massive theories \cite{Kraichnan}. 

It can be useful to rewrite the continuous family of massive scalar gravities given above using this exponential relationship; this definition of $\gamma$ is somewhat ecumenical.  
  The result is
\begin{equation}
\mathcal{L}_{ms} = \frac{ m^2 \sqrt{-\eta} }{64 \pi G} \frac{ [ w e^{8 \gamma \sqrt{\pi G}} - e^{8w \gamma \sqrt{\pi G}} + 1-w ]}{w(w-1)}; \end{equation}
the special cases $w=0$ and $w=1$  readily handled by l'H\^{o}pital's rule.
Using the series expansion for the exponential function, which converges everywhere, one has, reassuringly, a series involving quadratic (mass) and higher (self-interaction) terms:
\begin{eqnarray}
\mathcal{L}_{ms} = -\frac{ m^2 \sqrt{-\eta} }{64 \pi G}   \sum_{j=2}^{\infty}  \frac{ (8 \gamma \sqrt{\pi G})^j}{j!} \frac{1-w^{j-1} }{1-w} 
=   \nonumber \\  - \frac{m^2 \gamma^2 \sqrt{-\eta} }{2}  - \frac{4 m^2 \sqrt{\pi G} (1+w)  }{3} \gamma^3    +     \ldots.
\end{eqnarray}

 A given massive scalar gravity, such as for a specific value of $w$ from the family above, is either not the manifestation of any geometry, or is just one of many possible manifestations of the geometry comprised of two metrics, one flat and one conformally related to it. 
Thus a dynamics-first approach to relativity, or something like it, is appropriate \cite[Ch. 9 and Appendix A]{BrownPhysicalRelativity}---and one can see this from examples that could have been, and almost were, invented in the 1910s.  One could hardly claim that space-time's being flat, conformally flat, bimetric, or whatever it is in these massive scalar gravity theories, provides a ready explanation of the detailed phenomenology exhibited by the field equations, because, among other reasons, the geometry does not pick out a value of $w.$  In this respect, the recently derived infinity of scalar gravities  marks a \emph{conceptual} advance over the derivation of only one \cite{FreundNambu}, an advance that could be anticipated by consideration of the massive spin 2 analogs \cite{OP}.  Neither does the arguable existence of Minkowski space-time explain why rods and clocks exhibit Minkowskian behavior as Norton expects \cite{NortonFails}, because they \emph{don't} exhibit Minkowski behavior.  Rods and clocks, if they exhibit any unique geometry, exhibit the geometry to which matter couples; thus one can inspect  the matter Lagrangian and see that rods and clocks do not exhibit Minkowski geometry. As Norton elsewhere urges, the philosophy of geometry is not an enterprise rightly devoted to giving  a spurious air of necessity to whatever theory is presently our best \cite[pp. 848, 849]{Norton}.  So merely possible theories, especially those that are so similar to realistic physics and readily derived using well motivated principles,  should be entertained when assessing issues of the explanatory priority of laws \emph{vs.} geometry \cite[Ch. 9 and Appendix A]{BrownPhysicalRelativity}, or, for that matter, the conventionality \emph{vs.} empirical factuality of geometry.
Yet critiques of Brown's work routinely default to single-metric theories, or more specifically even to ``our best scientific theory.''   
Strikingly, a similar omission occurred in the 1970s critiques of Gr\"{u}nbaum's conventionality of geometry thesis, as will appear below.  In taking laws as prior to geometry, I do not mean to imply that laws are entirely free of geometrical presuppositions, but that geometry is adequately encoded in the laws (such as in the Lagrangian density) in any theory, but it in most theories it is not adequately described by a list of geometric objects.\footnote{I thank a referee for helping to formulate this point.}  Massive scalar gravity(s) show that this inadequacy extends even into what one might describe as Special Relativity or Minkowski space-time.


\section{Massive Scalar Gravity as Violating Einstein's Principles}

The mathematical-philosophical side of Einstein's process of discovery for General Relativity involved various Principles:  general relativity, general covariance, equivalence, and Mach. (For a brief useful history of the principles, including how Mach's principle was split off from the principle of general relativity, see (\cite{LehnerPrinciples}).)  
 Part of the interest of massive scalar gravity, like massive variants of General Relativity \cite{FMS}, is that though massive scalar gravities are perfectly sensible special  relativistic field theories, they violate all of Einstein's principles, at least in the robust senses in which Einstein used them to derive substantive conclusions.  Massive scalar gravities admit the inertial references frames of special relativity but no larger set of reference frames, contrary to the principle of general relativity.  Massive scalar gravities privilege the Poincar\'{e} transformations as relating preferred coordinate systems, while arbitrary coordinate systems are admissible only using tensor calculus to introduce fudge factors such as the metric \emph{tensor} and Christoffel symbols, contrary to the stronger and more interesting (as opposed to purely formal) sense of general covariance in which the metric tensor and Christoffel symbols don't enter simply to correct for failure to employ the privileged Cartesian coordinates.  Massive scalar gravities distinguish between gravitation and inertia, because inertia is characterized by the flat metric tensor, while the gravitational potential is a distinct scalar (density) field.  Thus the principle of equivalence (read strongly as the identification of gravity and inertia, or a bit more weakly as the claim that gravity and inertia have the same effects\footnote{I thank a referee for pointing out that this weaker claim also fails.}) fails for massive scalar gravity, though empirically the difference between gravity and inertia is \emph{observable} only in experiments sensitive to the graviton mass term and hence involving long distances.   In massive scalar gravity, contrary to Machian expectations, inertia has a core that is absolute and given from above, but gravity also modifies inertia somewhat; the core can be distinguished if one is looks carefully enough to observe the influence of the mass term in the gravitational field equation.  In short, massive scalar gravity shows how readily conceivable is the failure of most or all of Einstein's principles in the strong interesting senses, and hence how frail a justification for Einstein's equations must be if it relies on such principles as premises. 
A similar point has been made by Luc Blanchet \cite{BlanchetNonmetric}.  
 According to Robert DiSalle, 
\begin{quote}
Einstein thought that anyone who followed the philosophical steps that he had taken, whatever their [\emph{sic}] scientific background, would be convinced of the basic principles of special and general relativity.  By the later twentieth
century, however, philosophers came to think of those steps as somewhat
arbitrary, and as not very clearly related to the theories that Einstein actually
produced. They had a heuristic value for Einstein, and may have again
for a future theory of space-time. To believe again that such philosophical
arguments could be crucial---not only to the motivation for a theory, but
also to its real significance in our scientific understanding of the world---
we need a more philosophically subtle and historically realistic account of
those arguments, and the peculiar roles that philosophy and physics have
played in them. \cite[p. xi]{DiSalleSpacetime}
\end{quote}
 Norton on occasion has felt the need to try to dispel the mystical air surrounding many common justifications for Einstein's equations  by seeking eliminative inductions \cite{NortonEliminative}.

Violating Einstein's principles with a theory that was empirically falsified in 1919 due to the bending of light might seem to set the bar too low.  Who cares if a wildly obsolete and wrong theory violates Einstein's principles?  One should care because so much of what one learns about massive scalar (spin-$0$) gravity has a good chance of being true also about massive spin-$2$ (tensor) gravity also.  Hence one needs to consider massive spin-$2$ gravity and show that either it doesn't work, or that it isn't like massive scalar gravity in the given respect after all, or else one should anticipate violating the strong interesting versions of Einstein's principle with a currently adequate theory. 

By contrast Peter van Nieuwenhuizen could describe roughly what philosophers would call an eliminative induction leading to Einstein's theory.  Having moved beyond scalar theories (which do not bend light), the next option for gravity is a symmetric rank 2 tensor (spin 2).  
 Recalling that ``ghosts'' are negative-energy degrees of freedom, which are expected to produce instability, while tachyons move faster than light and hence should be excluded to preserve the usual relativistic notion of causality,  
\begin{quote}
[t]he conclusions are that the only tensor theories without ghosts or tachyons,
which contain spin two, are linear Einstein theory and [massive] Fierz-Pauli theory. Hence, the
gauge invariance and locality of gravitation follow from the absence of ghosts and
need no longer be postulated separately [reference to \cite{Sexl,NSSexlField,NSSexlLinear}], and general relativity follows from special
relativity by excluding ghosts [reference to \cite{Deser}]. Also it follows that any linear or non-linear
theory of gravitation has a discontinuous mass limit*. [footnote to \cite{DeserMass}]  \cite{VanN}. \end{quote} 
At a time when historians of General Relativity have made a  detailed study of Einstein's notebooks from \emph{c.} 1914  \cite{JanssenRenn,RennSauerPathways} 
that has shown the importance of his physical strategy involving energy-momentum conservation and the analogy to electromagnetism, it is striking to see both how similar the later particle physics work is in general outline and how much more compelling the later argumentation is. A key ingredient added by particle physicists is testing for ghosts.  It seems not to have been noticed that Einstein's \emph{Entwurf} theory is full of ghosts.  Even with that powerful new test, van Nieuwenhuizen's claim is a bit too quick:  besides massive spin-$2$ gravity, which seemed freshly falsified (at least empirically, if not \emph{a priori}) when he wrote but has experienced a revival since 2010, one could also include unimodular General Relativity (like General Relativity with a solution-dependent cosmological constant) and slightly bimetric theories \cite{SliBimGRG} (like scalar-tensor theories with a solution-dependent cosmological constant).  These theories are all closely related to Einstein's theory, however, so at least one arrives \emph{near} Einstein's theory in many respects, including nonlinear terms.   



Massive scalar gravities are not geometrizable in the usual sense: there is no way to absorb fully the gravitational potential into the geometry of space-time, at least if one does not permit multiple volume elements.  The mass term depends on both $\sqrt{-g}$ that contains the gravitational influence and $\sqrt{-\eta},$ which contains no gravitational influence.  Massive scalar gravity, like Seeliger-Neumann-Einstein `massive' Newtonian theory, makes a distinction between the space-time geometry and the gravitational field, albeit a subtle one.  By contrast Newtonian gravity in the geometrized Newton-Cartan form, Nordstr\"{o}m scalar gravity in geometrized Einstein-Fokker form, and General Relativity in its usual form combine the gravitational influence and space-time geometry such that neither gravity nor any supposed originally flat geometry appears separately.  If one permits more than one volume element in scalar gravity, then one can have any dependence whatsoever on the gravitational potential, making geometrization toothless and purely formal rather than substantive.
The same holds for multiple metrics, multiple connections, \emph{etc.}:  one could easily hide a gravitational potential (or its gradient) as the difference between them.    Massive scalar gravities contain both the conformally flat metric $g_{\mu\nu}=\hat{\eta}_{\mu\nu} \sqrt{-g}^\frac{1}{2}$  of Nordstr\"{o}m's theory (as geometrized by Einstein and Fokker \cite{EinsteinFokker} and improved with help from T. Y. Thomas) and the flat metric  $\eta_{\mu\nu}=\hat{\eta}_{\mu\nu} \sqrt{-\eta}^\frac{1}{2}$  of SR. (Massive Newtonian gravity, that is, Neumann-Seeliger-Einstein gravity, is not geometrizable either.)  Because there are two metrics present (albeit conformally related), one has a good argument for the conventionality of geometry, as Poincar\'{e} envisaged (on which more below). For the same reasons, strong versions of the equivalence principle are not admissible; gravity is not a feature of space-time geometry, though it looks that way unless one makes sufficiently precise measurements involving cosmic distances due to the smallness of the graviton mass.   It is only the empirical fact of the bending of light by gravity, not any inherent conceptual defect, that made it impossible to treat gravity adequately as a special relativistic theory of a massive scalar field (\emph{c.f.} \cite{GiuliniAbsolute,GiuliniScalar}).   Had Nordstr\"{o}m's theory still  been viable by the time that Wigner's classification of Lorentz group representations in terms of mass and spin was widely known, it seems certain that massive scalar gravity would have been considered.  Its neglect until 1968 \cite{FreundNambu,DelbourgoSalamStrathdee,DeserHalpern}, if not the present, is one of the many disadvantages from the well known gulf \cite{Rovelli,Feynman} between general relativists and particle physicists.  The precedent that should have been noticed for massive scalar gravity suggests by analogy that one could consider massive tensor gravity as well.  


\section{Conventionalism \emph{vs.}  Empiricism, Ehlers-Pirani-Schild, and Poincar\'{e}'s Modal Argument}

Mathematicians have recently found it worthwhile to study what one can do with a metric and an additional volume element, under the name ``metric-measure space''  \cite{LottMetricMeasure}.  Such a framework fits the well known Brans-Dicke scalar-tensor theory and various related theories \cite{BransDicke}, hence is not very new.  In fact such possibilities, in the simplest cases, are more than a century old.  Poincar\'{e}  envisaged the possibility of a theory in which more than one metric played a role   and saw this possibility as motivating the conventionality of geometry \cite[pp. 88, 89]{PoincareFoundations} \cite{BenMenahemPoincare}. 
Let us call this ``Poincar\'{e}'s modal argument for the conventionality of geometry.''  To assess this argument, one needs to be a bit clearer than usual about the main competition of the same era, empiricism.  In important respects the old empiricism \emph{vs.} conventionalism debate of the 1920s (such as Eddington \emph{vs.} Poincar\'{e} \cite{EddingtonSTG}) has been recapitulated by the realism \emph{vs.} conventionalism debate of the 1970s (such as Putnam, Stein, Earman and Friedman \emph{vs.} Gr\"{u}nbaum) and the recent realism \emph{vs.} constructivism debate \cite{NortonFails,BrownPhysicalRelativity}.  In all three cases a more or less unrecognized issue leading the two sides to talk past each other was a \emph{disagreement about modal scope}, with the conventionalist/constructivist side envisaging a broad modal scope and hence discussing alternative theories, and the empiricist/realist side considering primarily `our best theory,' General Relativity.  Given that the two sides (if the reader will accept my amalgamation of conventionalism and constructivism, and of empiricism and realism, for present purposes) have asked different questions, it becomes less surprising that the answers differed, and more plausible that both views contain important insights.  
Despite criticisms of conventionalism \cite{NortonConvention} and constructivism \cite{NortonFails}, Norton has urged (as noted above) that the philosophy of geometry is not an enterprise rightly devoted to giving  a spurious air of necessity to whatever theory is presently our best \cite[pp. 848, 849]{Norton}.


Empiricism comes in two different senses relevant to geometry.  The broader sense involves the revisability in principle of even deeply entrenched and intuitively plausible ideas  and the entertainment of as many options as possible in light of experience, as opposed to apriorism. Empiricism in this sense is, in my view, a good idea.  Geometric empiricism, by contrast, involves the claim that the geometry of space(time) can and should be ascertained empirically, but it came to mean something stronger.  In Helmholtz's time, geometric empiricism was still an instance of empiricism in the broader sense.  But the progress of mathematics from  1898 made a flat metric together with conformally related curved metric (agreeing on angles but not volumes) available  \cite{CottonCR2,Cotton,PoincareFoundations,Finzi,Fubini1905,WeylReineInfinitesimal,SchoutenStruikBimetric,SchoutenConformal,Finzi1,Finzi2,KasnerLight,KasnerSolar,BrinkmannConformal,StruikDG,SchoutenRK,LeviCivita,SchoutenStruik},
 and indeed fairly common in 3 dimensions, in French, Italian and German, and eventually in English.  Thus Poincar\'{e} the philosopher-mathematician was aware of the possibility of there being no specific fact of the matter about geometry  \cite[pp. 88, 89]{PoincareFoundations} \cite{BenMenahemPoincare}.  Just this possibility, upgraded from space to space-time, is realized physically  massive scalar gravities.  The geometric empiricism of Eddington \cite[p. 10]{EddingtonSTG} \cite[pp. 159-162]{EddingtonNature}, by contrast, had hardened into a dogma.  It did so by presupposing that there exists a unique geometry---long after alternatives to that presupposition were available, as even his own work on affine gravity suggests \cite{EddingtonAffine,GoennerAffineEddington}.  Later geometric empiricism thus rejected \emph{a priori} such theories as might not have a unique geometry.  The tendency is to freeze the development of mathematics and of the physics that employs it in the 1910s. By stopping history early, one can make General Relativity appear as the end of it.

 Whereas physicists aware of the particle physics tradition often incline toward conventionalism \cite{Feynman,FMS,SexlConvention,Weinberg} and sometimes employ multiple metrics due to a graviton mass in particular \cite{OP,FMS},  philosophical works aiming to refute conventionalism tend not to address  particle physics or multi-geometry theories \cite{EarmanCongruence,FriedmanGrunbaum,PutnamConventionLong,SpirtesConvention,FriedmanFoundations,Torretti,ColemanKorteHarmonic,NortonConvention}. (A partial exception is (\cite{SteinGrunbaum}).)  From the broad modal perspective, single-geometry theories such as Newtonian physics, Special Relativity, Nordstr\"{o}m's theory, and General Relativity, are a special and somewhat degenerate case, albeit one of great importance as well as convenience.  Single-geometry theories are the home turf of geometric empiricism (and its later cousin realism).  In that context the massless spin $2$ particle physics derivations  \cite{Kraichnan,Gupta,Feynman,FMS,Deser,SliBimGRG,BenMenahemConventionalism} achieve rough parity for conventionalism \emph{via} something like a demonstrative and eliminative induction for Einstein's equations starting from a field theory in flat space-time.  Conventionalism about geometry indeed would be disappointing if universal forces were the conventionalist's only or main idea, as sometimes is suggested.  But on the contrary, universal forces are rather the response to the \emph{hardest} cases for conventionalism, cases where \emph{it is meaningful} to talk about \emph{the} geometry of space(time). 
Thus Norton's critique of Reichenbach's conventionalism and universal forces speaks of ``the [metric] revealed by direct measurement'' \cite[159]{NortonConvention} and ``the [metric] revealed by uncorrected distance measurements.'' \cite[p. 166]{NortonConvention}  But why must there be any such thing?  Reichenbach \cite[appendix only in the German original]{ReichenbachRaum}\footnote{I thank Marco Giovanelli for pointing me to the newly available draft translation \cite{ReichenbachAppendix} of this long lost appendix.} and Gr\"{u}nbaum (below) had already seen in metric-affine theories the possible nonexistence of such a thing.


The difficulty can now be seen in the longstanding and sophisticated Coleman-Kort\'{e} argument for the non-conventionality of geometry \cite{ColemanKorteJet,ColemanKortePSA,ColemanKorteConstraints,ColemanKorteSemantic1,ColemanKorteSemantic2,KorteWeylScholz,WeylKorteStanford}  from the Ehlers-Pirani-Schild (EPS) construction.  Purged of chaff \emph{via} attention to irreducible geometric objects, the EPS construction is a charming exercise in differential geometry, relating the $36$ components of the symmetric covariant derivative of a (unimodular) conformal metric density to the $36$ components of a (traceless) symmetric projective connection \cite{WeylConformalProjective,Schouten,EhlersPiraniSchild,KorteWeylScholz}.  (Note that a projective connection is not a connection, but a weaker structure that makes sense in its own right.)  If one makes the volume connection integrable, then one gets a metric up to a constant numerical factor.  This presentation highlights the utility of irreducible geometric objects (conformal metric density, volume element, projective connection, volume connection) and the utility of classical differential geometry for isolating them.  Unfortunately the anti-conventionalist argument hinges entirely on the assumption that there is exactly one physically relevant conformal metric density and exactly one physically relevant projective connection.  If one has more than one conformal metric density, or more than one projective connection, then one can run the Ehlers-Pirani-Schild construction multiple times!  Early critics of anti-conventionalist arguments from the EPS construction realized that it requires a conventional choice, such as a choice of inertially moving bodies \cite[ch. 22]{GrunbaumSpace} \cite[p. 197]{WinnieConvention} \cite[p. 259]{SklarFactsEarman} \cite[p. 295]{SalmonConvention}.  A choice of inertially moving bodies is in effect a choice of projective connection.  
Perhaps the conventionalists did not emphasize enough (though the themes can be found) that (1) the most interesting question is the modally broad one that considers a variety of theories rather than a single theory (even `our best'), and (2)  there exists a moderately interesting physical theory (or many of them) containing within itself `rival' geometries, that is, more than one example of a given type of geometric object.  Thus EPS manifestly fails to undermine conventionality for any theory with  multiple geometries---the sort that made conventionalism especially attractive anyway.

As Ben-Menahem pointed out in modern times \cite{BenMenahemPoincare},
Poincar\'{e} already envisaged the possibility of theories in which some kinds of matter exhibit one geometry, but other kinds of matter see another geometry. 
He writes:
\begin{quote}
Suppose, for example, that we have a great sphere of radius $R$ and that the temperature decreases from the center to the surface of this sphere according to the law of which I have spoken in describing the non-Euclidean world.

We might have bodies whose expansion would be negligible and which would act like ordinary rigid solids; and, on the other hand, bodies very dilatable and which would act like non-Euclidean solids.  We might have two double pyramids $OABCDEFGH$ and $O^{\prime}A^{\prime}B^{\prime}C^{\prime}D^{\prime}E^{\prime}F^{\prime}G^{\prime}H^{\prime}$ and two triangles $\alpha\beta\gamma$ and $\alpha^{\prime}\beta^{\prime}\gamma^{\prime}$. The first double pyramid might be rectilinear and the second curvilinear; the triangle $\alpha\beta\gamma$  might be made of inexpansible matter and the other of a very dilatable matter.

It would then be possible to make the first observations with the double pyramid $OAH$ and the triangle $\alpha\beta\gamma$, and the second with the double pyramid   $O^{\prime}A^{\prime}H^{\prime}$ and the triangle $\alpha^{\prime}\beta^{\prime}\gamma^{\prime}$.  And then experiment would seem to prove first that the Euclidean geometry is true and then that it is false.

\emph{Experiments therefore have a bearing, not on space, but on bodies.}  \cite[pp. 88, 89]{PoincareFoundations} (emphasis in the original) 
\end{quote}
One might update the last sentence to say that experiments have a bearing not on the geometry of space-time, but on the way that geometry(s) appear in the Lagrangian densities for the various matter fields.

  Ben-Menahem, who might be the first to make much of this passage at least for a long time, comments:
\begin{quote}
Poincar\'{e} goes to great lengths to show that it is conceivable that different types of objects conform to different geometries.  We could ask a mechanic, he says, to construct an object that moves in conformity with non-Euclidean geometry, while other objects retain their Euclidean movement.  In the same way, in his hypothetical world, bodies with negligible contraction, that behave like ordinary invariable solids, could coexist with more variable bodies that behave in non-Euclidean ways. [Quotation of Poincar\'{e} suppressed to avoid repetition.]  Is it absurd, according to Poincar\'{e}, to relinquish the quest for a unified geometry?  Probably, on pragmatic grounds; but it is not incoherent.  The conceivability of such pluralism is another point in favor of conventionalism.    \cite[p. 489]{BenMenahemPoincare} 
\end{quote}
Apart from massive gravities, the multiple-metric option resurfaced in physics around 1960 with scalar-tensor theories due to Brans and Dicke, with some 1950s ideas by Jordan and Thiry involving a higher space-time dimensionality.  Philosophers eventually noticed this and other examples and drew conventionalist and/or constructivist conclusions \cite{WeinsteinScalar,BrownPhysicalRelativity}, in effect reinventing Poincar\'{e}'s argument in light of now-extant examples.
A fresh physics paper asks a good question in its title article:  ``The Nature of Spacetime in Bigravity: Two Metrics or None?'' \cite{SolomonBigravityFinsler}  Such questions could and perhaps should have appeared in a specific physical theory the 1910s or 1920s with the proposal of massive scalar gravity.  How does the later hardened geometric empiricism or the new realism address such possibilities?

Massive scalar gravity, had it been available to Reichenbach, would have diverted attention at least partly away from universal forces, toward  \emph{almost-universal} forces. In massive scalar gravities, matter seems a conformally flat metric, the volume element being a combination of a non-dynamical part $\sqrt{-\eta}$ and the gravitational potential $\gamma$, a literal instantiation of the Poincar\'{e} epistemological sum of geometry and physics.  But the dynamics of gravity itself exhibits the full flat background geometry.  Thus one cannot set gravity, the almost universal force, to $0$.  (Recall that such annihilation of universal forces was hailed by Carnap as a great insight of Reichenbach's \cite[p. vii]{ReichenbachSpace}.)  And yet one can never be sure, on account of the smoothness of the massless limit, that the graviton mass is zero; one has permanent underdetermination from approximate but arbitrarily close empirical equivalence \cite{UnderdeterminationPhoton}.  Thus one can never be sure that the observable geometry is the same as the geometry pertaining to the symmetries of the theory's laws.  Such a result would hold, at least \emph{prima facie}, also for massive variants of General Relativity, the conception of which is greatly facilitated by the spin 0-spin 2 analogy, replacing a scalar by a symmetric tensor.

 But the deep point of conventionalism about geometry is not that the true geometry is flat, or even that one can retain flat geometry. Though that point is less implausible than many have thought since the rise of General Relativity \cite{FMS}, it doesn't even apply for Brans-Dicke scalar-tensor gravity \cite{WeinsteinScalar,FaraoniNadeau}.  The deep point is rather that questions about the `true geometry' have no good answers in general, do not need good answers, and are not very interesting. It is precisely the dearth of interesting theories that can be built out of just a single metric that makes geometry so informative in the case of General Relativity (massless spin 2) and Nordstr\"{o}m's theory (massless spin 0).  Hence the need to expand the modal scope by investigating a larger class of well motivated theories arises, especially theories without a unique (single-metric) geometry.

\section{Gr\"{u}nbaum on Riemann's Concordance Assumption, Putnam}

The proposal that conventionalists/constructivists and empiricists/realists have been talking past each other due to different assumed modal scope helps to shed a little light on some notoriously opaque debates regarding conventionalism in the 1960s-70s.  Earman, broadly sympathetic to Putnam's critique of Gr\"{u}nbaum on conventionality but differing on some details, provides a useful example of focusing attention on the `best theory'; Gr\"{u}nbaum may not always have presented conventionalism in (what I take to be) its strongest light.  Earman wrote:
\begin{quote} \ldots my goal is the limited one of showing that (i) contrary to
Gruenbaum, the general theory of relativity (hereafter, GTR) does not
support the claim that there is a latitude for a conventional choice in the
standards of spatial and temporal congruence and that (ii) Gruenbaum's
thesis of the metrical amorphousness of space and time does not illuminate
and does not draw support from the GTR. \cite{EarmanCongruence}  
\end{quote}  If single-geometry theories are a special case, one where empiricism and realism flourish and conventionalism and constructivism seem unhelpful, then Earman's points fit perfectly.  One \emph{could} define parallel transport and hence congruence in a way incompatible with $g_{\mu\nu}$, but doing so would be contraindicated by the convenience, empirical guidance, and judgment forming parts of Poincar\'{e}'s conventionalism at least.  Earman's immediate deprecation of the result also fits:
``\ldots I do not view the achievement
of this goal as being terribly important in itself\ldots'' \cite{EarmanCongruence}.  The crucial next question is what, if anything, one says about the broader modal scope, theories that, unlike General Relativity, have multiple geometries.  These theories, though individually unimpressive in most cases, are far more numerous than single-geometry theories and could be true.  They have been taken with much greater seriousness in the past five years (massive spin-$2$ gravity, bigravity) than at most times in the past.

Something like Poincar\'{e}'s modal argument for conventionality is suggested by Gr\"{u}nbaum's invocation of Max von Laue's discussion of bimetric theories \cite[note 74]{GrunbaumEarman}  \cite[pp. 186-196]{LaueBook}, which mentions work by Levi-Civita, Nathan Rosen, 
 Achille Papapetrou, and Max Kohler.
Because a space-time metric induces a connection, two metrics induce (at least) two connections.  Actually one can define more than two connections, most obviously by splitting metrics into conformal and volume pieces, splitting connections into projective (traceless) and pure trace (volume) pieces, and mixing and matching. There seems to be no objection to taking arbitrarily weighted geometric means of the volume elements $ \sqrt{-g}^u \sqrt{-\eta}^{(1-u)}$ 
(giving arbitrarily weighted arithmetic means of the volume connections by logarithmic differentiation) and arbitrarily weighted arithmetic means of the projective connections $v\hat{\Gamma}^{\alpha}_{\mu\nu} + (1-v)\hat{\{ ^{\alpha}_{\mu\nu}  \} }$, achieving a 2-parameter family of connections using just two metrics.  
 Hence bimetric geometry from the 1920s raises most or all the issues raised by metric-incompatible connections from the late 1910s.\footnote{A classic treatment of the non-metricity tensor is Schouten's  \cite{Schouten}.}  Clearly the connection associated with one metric will generically be incompatible with a different metric (excepting the trivial case where one metric is a constant times the other).  Of course there is no direct route from having some geometrical entity in a theory to that entity's having direct chronogeometric significance \cite{FMS,BrownPhysicalRelativity,ButterfieldCausalityConventionGeometry,KnoxEffectiveTorsion}.  One has to look at the matter Lagrangian density to ascertain what geometry(s) matter actually sees.  It also isn't terribly easy to get every type of matter to couple to a connection except \emph{via} a metric.  (The ``hypermomentum'' of Friedrich Hehl and collaborators would fill in this gap \cite{HehlGauge}.)  Hence in considering  stories about the physical meaning of a connection incompatible with the/a metric,\footnote{There is of course  a large difference between a connection's being incompatible with some metric, which is a relation between that connection and that metric, and a connection's not being compatible with \emph{any} metric.  Being compatible with some specific metric yields a familiar but optional  antisymmetry of the Riemann tensor with one index lowered by that metric \cite[p. 39]{Wald} \cite[pp. 324, 325]{MTW}. For more detail see (\cite{Edgar1,Edgar2,EdgarWeyl}).   An even stronger sort of metric-incompatibility comes from not being compatible even with any volume element; such a property (not a relation) implies that the trace of the connection coefficients has non-zero curl, which is just the non-vanishing of the other trace of the primordial $(1,3)$ Riemann tensor.} one wants to think in some detail about their physical realizability.

The idea(s) that parallel transport might fail to be compatible with (the? a?) metric is an interesting issue highlighted by Gr\"{u}nbaum, quite separable from the question of the clarity of his distinction between intrinsic and extrinsic metrics. I take the liberty of making a lengthy quotation, which is punctuated by useful quotations from Einstein, Marzke and Wheeler, and Reichenbach.  
\begin{quote}  
1. Let us note first that an important \emph{empirical} hypothesis actually underlies the
alleged conceptual necessity to metrize any kind of $P$-space by means of rods and
tapes. The pertinent hypothesis was characterized as empirical by Einstein as
follows:
\begin{quote} 
All practical geometry is based upon a principle which is accessible to experience, and
which we will now try to realise. We will call that which is enclosed between two boundaries,
marked upon a practically-rigid body, a tract. We imagine two practically-rigid bodies,
each with a tract marked out on it. These two tracts are said to be ``equal to one another'' if
the boundaries of the one tract can be brought to coincide permanently with the boundaries
of the other. We now assume that:

If two tracts are found to be equal once and anywhere, they are equal always and everywhere.

Not only the practical geometry of Euclid, but also its nearest generalisation, the practical
geometry of Riemann and therewith the general theory of relativity, rest upon this assumption
([reference to a reprint of \cite{EinsteinGeomExp}]). \end{quote}
As I have done previously elsewhere ([\cite{GrunbaumGeometry}], pp. 272, 277), I shall refer to the empirical
assumption just formulated by Einstein as ``Riemann's concordance assumption,''
or, briefly, as ``RCA.''\ldots 

We see that the empirical truth of RCA plays the following role: It is a necessary
condition for the \emph{consistent} use of rigid rods in assigning lengths to space intervals
that any collection of two or more initially coinciding unit solid rods of whatever
chemical constitution can thereafter be used \emph{interchangeably} everywhere in the
$P$-manifold \emph{independently of their paths of transport, unless} they are subjected to
independently designatable perturbing influences. Thus, the assumption is made
here that there is a concordance in the coincidence behavior of solid rods such that
no inconsistency would result from the subsequent interchangeable use of initially
coinciding unit rods, if they remain \emph{unperturbed} or ``rigid'' in the specified sense.
In short, there is concordance among rigid rods such that all rigid unit rods alike
yield the same metric and the same geometry. It will be recalled that in section 2(c),
(i), we had occasion to cite the following comment on RCA by Marzke and
Wheeler:
\begin{quote} 
This postulate is not obvious and, in principle, could even be wrong. For example,
Weyl once proposed (and later had to give up) a unified theory of electromagnetism and
gravitation in which the Riemann postulate was abandoned. In Weyl's theory, two measuring
rods, cut to have identical lengths at a point $A$ in space-time, and carried by different
routes to a point $C$, will \emph{differ} in length when they are brought together ([\cite{WheelerMarzke}], p. 58). \end{quote} 
There we also noted that Marzke and Wheeler consider ``what kind of physics
would not be compatible with Riemann's postulate,'' offer the ``Validity of Pauli
Principle as Partial Evidence for Riemann's Postulate'' ([\cite{WheelerMarzke}], p. 60), but conclude
([\cite{WheelerMarzke}], p. 61) that ``It would be desirable to have a more \emph{decisive} experimental argument
for the Riemannian postulate.''

In the philosophical literature, Reichenbach has given a vivid description of the
kinds of phenomena that would occur if solid bodies were to \emph{violate} RCA, as they
do in Weyl's kind of nonRiemannian geometry. This description, which appears in
the German original of his classic \emph{Philosophie der Raum-Zeit-Lehre} but not in the
1958 English translation [58], runs as follows:
\begin{quote} 
\ldots whether we could put, say, six chairs in a row into a room would depend on the path
by which the chairs were to be brought into the room, and we might perhaps first have to
let our chairs make a trip around the world so that the room could accommodate them. By
the same token, it would be uncertain whether a visitor could fit onto one of the chairs;
this would depend on his prior trajectory. Such states of affairs might perhaps strike us as
strange, but they are logically possible; and were they to obtain, humans would surely
have come to terms with them ([57], p. 333; translation is mine [Gr\"{u}nbaum's]).  
\end{quote} 
It is, of course, irrelevant to the issue posed by Swinburne that Weyl's nonRiemannian
geometry, in which rods are held to violate RCA, did not succeed empirically
as a unified theory of electromagnetism and gravitation. Instead, the import of the
logical possibility that the hypothesis RCA might be empirically false emerges
from the following considerations.\ldots 
\cite[pp. 571, 572]{GrunbaumSpaceTimeFalsifiability}  
\end{quote}

Where Gr\"{u}nbaum finds a significant philosophical issue (more or less correctly, though my caution about achieving chronogeometric significance for a geometrical structure in the physical laws \emph{via} a Lagrangian density, rather than by `by hand', should be recalled), and it involves a broad modal scope, Putnam only finds paradox:
\begin{quote}
Reichenbach used to begin his lectures on the Philosophy of Space and Time in a way which already brought an air of paradox to the subject. He would take two objects of markedly different size, say an ash tray and a table, situated in different parts of the room, and ask the students ``How do you know that one is bigger than the other?''  \cite{PutnamConventionLong}
\end{quote}  
Apparently such possibilities were not among those that Putnam was prepared to conceive; the problem is Putnam's.  Insofar as one can think of a rigid rod as a vector to undergo parallel transport, the possibility of parallel transport according to a connection not derived from or even compatible with\footnote{Torsion offers the possibility of a connection compatible with a metric but not derived from it \cite{KnoxEffectiveTorsion}.} the metric that one is considering can be readily entertained; it is part of the great conceptual advance of recognizing the connection as an entity independent of a metric \cite{Schouten}.  One might think that parallel transport with respect to a metric-incompatible connection is always merely due to an unwise choice of connection.  But there are theories in which non-metricity plays a dynamical role \cite{Hehl,HehlVary2}, which one ought not to refuse to entertain.  Furthermore, there are connections that are not compatible with \emph{any} metric. Hence simply making a wiser choice of connection to avoid non-metricity might not be an option.  Reichenbach and Gr\"{u}nbaum both explored the idea of connections not derived from a metric \cite[untranslated German appendix]{ReichenbachRaum} \cite{ReichenbachAppendix,Reichenbach1929Zeitschrift,ReichenbachAngewandte,GrunbaumEmpty}.  By contrast, Putnam's work on space-time physics \cite{PutnamGrunbaum,Putnam,PutnamConventionLong} seems not to profit from developments in physics and geometry after 1916, such as Weyl's work on connections that are not based wholly on a metric, not to mention Levi-Civita's bimetric geometry or the whole tradition of particle physicists' work on gravity.  Even after Gr\"{u}nbaum's detailed consideration of the possible failure of Riemann's concordance hypothesis \cite{GrunbaumSpaceTimeFalsifiability}, Putnam's discussion defaults to single-metric General Relativity, evading the issue of what to make of theories with multiple or `rival' geometric ingredients.  
According to Putnam's early critique of Gr\"{u}nbaum's philosophy of geometry, 
\begin{quote}  In sum: our alleged `freedom' to \emph{choose a different $g_{ik}$ tensor} (a different space-time metric)
\emph{at the cost of complicating the laws of nature is in fact never employed in the
general theory of relativity}. All observers are required to `choose' the \emph{same} space-time metric.
\cite{PutnamGrunbaum} \end{quote}  
Putnam has defaulted to the narrow scope of `our best theory' and so has surreptitiously achieved a sort of home-field advantage.  But still he falters:  one can perhaps rescue this descriptive claim from effortless falsification even using literature before 1963 (including Gupta's foundational work on quantum gravity \cite{Rosen1,Rosen2,Papapetrou,Kohler,Kohler2,LaueBook,GuptaPPSL2,Gupta,GuptaReview}) if one carefully gerrymanders the physics literature so that the many counterexamples, some of them by eminent authors, do not count as ``the general theory of relativity.'' Putnam wrote those words at roughly the time of Feynman's significant further development of particle physics treatments of General Relativity \cite{Feynman63}, which circulated informally \cite{FeynmanUnpublished} and then eventually mostly were published as a book \cite{Feynman}.
 But then one will need to credit other authors with inventing an alternative theory that shares the same field equations, or uses the same field equations as a classical starting point before quantizing, or the like.  Space-time physics is a bit like heaven and earth as portrayed by Hamlet  to Horatio:
\begin{quote}  ``[t]here are more things in heaven and earth\ldots \\ Than are dreamt of in your philosophy.'' \end{quote}

Subsequent work will discuss how massive scalar gravity, if entertained at the right time in the right context, would have blocked Moritz Schlick's sociologically successful overthrow of Kantian synthetic \emph{a priori} knowledge, and how Einstein's widely followed false analogy between his reinvented Seeliger-Neumann modification of gravity and his cosmological constant \cite{Schucking} facilitated the neglect of massive (spin 0 and spin 2) gravity during 1917-1939, much of the era when the fate of Kantian philosophy was settled negatively and logical empiricism took shape.  



\section{Appendix:  Malament-Weatherall-Manchak Conformal Restriction of Conventionality Evaded}

It has been argued that the conventionality of space-time metric geometry, whatever one makes of it, should be restricted to conformally related metrics, the null cone structure being factual rather than conventional  \cite{MalamentRotation,WeatherallManchakConventionality}.
That claim appears to be a technical point motivated by a conceptual point that causality is not conventional for Reichenbach.  Indeed causality is not conventional for Reichenbach, but the geometrical use of that null cone field that marks out causality, rather than an arbitrarily chosen one, \emph{is} conventional---at least, no less conventional than the metric is.  The geometry of angles, being a part of the geometry of distances, automatically inherits whatever conventionality there might be in metric geometry that survives the excision of volumes. 
Thomas's decomposition shows how to apply Reichenbach's conventionality-of-geometry equation $$ g'_{\mu\nu} +  F_{\mu\nu} = g_{\mu\nu}$$ \cite{ReichenbachSpace}  to the conformal part of a metric, \emph{pace} the claim  that the two metrics must be conformally related. 

 Using the Thomas-style decomposition for both $g'_{\mu\nu} $ and $g_{\mu\nu}$, and letting bars $| |$ signify a matrix determinant, one can show that
\begin{eqnarray}  \frac{ \hat{g}'_{\mu\nu} +  F_{\mu\nu} (-g')^{-\frac{1}{4}} }{(-|\hat{g}'_{\alpha\beta} + F_{\alpha\beta} (-g')^{-\frac{1}{4}}|)^{\frac{1}{4}} } = \hat{g}_{\mu\nu}. \end{eqnarray}
(The exponents $\pm \frac{1}{4}$ reflect the $4$-dimensionality of space-time, but the expression generalizes.)
Thus the universal `force' (potential) making the conformal geometry no less conventional than the metric geometry is $ F_{\alpha\beta} (-g')^{-\frac{1}{4}}$---or, better yet, a certain 90\% of it. At least if  $F_{\mu\nu}$ is small enough for perturbative expansion, then the universal force for the conformal geometry is the $\hat{g}'_{\mu\nu}$-traceless part of  $ F_{\alpha\beta} (-g')^{-\frac{1}{4}}$. One easily sees that if $F_{\mu\nu}$ is proportional to $ \hat{g}'_{\mu\nu},$ then it cancels out of the conformal part.  Other parametrizations of a metric and, concomitantly, its conformal part are of course possible \cite{IshamSalamStrathdeeExponential}.  
It would be an interesting but nontrivial task to say something global about these issues.

%

\section{Acknowledgments}
Thanks to Jeremy  Butterfield, Huw Price, John Barrow, Harvey Brown, Oliver Pooley, Don Howard and 
Katherine Brading for encouragement, Marco Giovanelli for finding the Reichenbach appendix, Nic Teh for discussion, and the referees for very helpful comments.  
This work was funded by the John Templeton Foundation grant \#38761.



\end{document}